\documentclass[twocolumn]{llncs}

\makeatletter
\@ifundefined{lhs2tex.lhs2tex.sty.read}%
  {\@namedef{lhs2tex.lhs2tex.sty.read}{}%
   \newcommand\SkipToFmtEnd{}%
   \newcommand\EndFmtInput{}%
   \long\def\SkipToFmtEnd#1\EndFmtInput{}%
  }\SkipToFmtEnd

\newcommand\ReadOnlyOnce[1]{\@ifundefined{#1}{\@namedef{#1}{}}\SkipToFmtEnd}
\usepackage{amstext}
\usepackage{amssymb}
\usepackage{stmaryrd}
\DeclareFontFamily{OT1}{cmtex}{}
\DeclareFontShape{OT1}{cmtex}{m}{n}
  {<5><6><7><8>cmtex8
   <9>cmtex9
   <10><10.95><12><14.4><17.28><20.74><24.88>cmtex10}{}
\DeclareFontShape{OT1}{cmtex}{m}{it}
  {<-> ssub * cmtt/m/it}{}

\DeclareFontShape{OT1}{cmtt}{bx}{n}
  {<5><6><7><8>cmtt8
   <9>cmbtt9
   <10><10.95><12><14.4><17.28><20.74><24.88>cmbtt10}{}
\DeclareFontShape{OT1}{cmtex}{bx}{n}
  {<-> ssub * cmtt/bx/n}{}

\newcommand{\Conid}[1]{\mathit{#1}}
\newcommand{\Varid}[1]{\mathit{#1}}
\newcommand{\anonymous}{\kern0.06em \vbox{\hrule\@width.5em}}

\usepackage{polytable}

\@ifundefined{mathindent}%
  {\newdimen\mathindent\mathindent\leftmargini}%
  {}%

\def\resethooks{%
  \global\let\SaveRestoreHook\empty
  \global\let\ColumnHook\empty}
\newcommand*{\savecolumns}[1][default]%
  {\g@addto@macro\SaveRestoreHook{\savecolumns[#1]}}
\newcommand*{\restorecolumns}[1][default]%
  {\g@addto@macro\SaveRestoreHook{\restorecolumns[#1]}}
\newcommand*{\aligncolumn}[2]%
  {\g@addto@macro\ColumnHook{\column{#1}{#2}}}

\resethooks

\newcommand{\onelinecommentchars}{\quad-{}- }
\newcommand{\commentbeginchars}{\enskip\{-}
\newcommand{\commentendchars}{-\}\enskip}

\newcommand{\visiblecomments}{%
  \let\onelinecomment=\onelinecommentchars
  \let\commentbegin=\commentbeginchars
  \let\commentend=\commentendchars}

\newcommand{\invisiblecomments}{%
  \let\onelinecomment=\empty
  \let\commentbegin=\empty
  \let\commentend=\empty}

\visiblecomments

\newlength{\blanklineskip}
\newcommand{\hsindent}[1]{\quad}
\let\hspre\empty
\let\hspost\empty

\EndFmtInput
\makeatother
\ReadOnlyOnce{polycode.fmt}%
\makeatletter

\newcommand{\hsnewpar}[1]%
  {{\parskip=0pt\parindent=0pt\par\vskip #1\noindent}}

\newcommand{\hscodestyle}{}

\newcommand{\sethscode}[1]%
  {\expandafter\let\expandafter\hscode\csname #1\endcsname
   \expandafter\let\expandafter\endhscode\csname end#1\endcsname}

  {\par\noindent
   \advance\leftskip\mathindent
   \hscodestyle
   \let\\=\@normalcr
   \let\hspre\(\let\hspost\)%
   \pboxed}%
  {\endpboxed\)%
   \par\noindent
   \ignorespacesafterend}

  {\hsnewpar\abovedisplayskip
   \advance\leftskip\mathindent
   \hscodestyle
   \let\hspre\(\let\hspost\)%
   \pboxed}%
  {\endpboxed%
   \hsnewpar\belowdisplayskip
   \ignorespacesafterend}

  {\hsnewpar\abovedisplayskip
   \advance\leftskip\mathindent
   \hscodestyle
   \let\\=\@normalcr
   \(\pboxed}%
  {\endpboxed\)%
   \hsnewpar\belowdisplayskip
   \ignorespacesafterend}

\newcommand{\plainhs}{\sethscode{plainhscode}}

\plainhs

  {\hsnewpar\abovedisplayskip
   \advance\leftskip\mathindent
   \hscodestyle
   \let\\=\@normalcr
   \(\parray}%
  {\endparray\)%
   \hsnewpar\belowdisplayskip
   \ignorespacesafterend}

  {\parray}{\endparray}

  {\(\parray}{\endparray\)}

\def\codeframewidth{\arrayrulewidth}
\RequirePackage{calc}

  {\parskip=\abovedisplayskip\par\noindent
   \hscodestyle
   \arrayrulewidth=\codeframewidth
   \tabular{@{}|p{\linewidth-2\arraycolsep-2\arrayrulewidth-2pt}|@{}}%
   \hline\framedhslinecorrect\\{-1.5ex}%
   \let\endoflinesave=\\
   \let\\=\@normalcr
   \(\pboxed}%
  {\endpboxed\)%
   \framedhslinecorrect\endoflinesave{.5ex}\hline
   \endtabular
   \parskip=\belowdisplayskip\par\noindent
   \ignorespacesafterend}

\newcommand{\framedhslinecorrect}[2]%
  {#1[#2]}

  {\(\def\column##1##2{}%
   \let\>\undefined\let\<\undefined\let\\\undefined
   \newcommand\>[1][]{}\newcommand\<[1][]{}\newcommand\\[1][]{}%
   \def\fromto##1##2##3{##3}%
   }{\) }%

  {\let\orighscode=\hscode
   \let\origendhscode=\endhscode
   \def\endhscode{\def\hscode{\endgroup\def\@currenvir{hscode}\\}\begingroup}
   \orighscode\def\hscode{\endgroup\def\@currenvir{hscode}}}%
  {\origendhscode
   \global\let\hscode=\orighscode
   \global\let\endhscode=\origendhscode}%

\makeatother
\EndFmtInput
\ReadOnlyOnce{agda.fmt}%

\RequirePackage[T1]{fontenc}
\RequirePackage[utf8x]{inputenc}
\RequirePackage{ucs}
\RequirePackage{amsfonts}

\DeclareUnicodeCharacter{737}{\textsuperscript{l}}
\DeclareUnicodeCharacter{8718}{\ensuremath{\blacksquare}}
\DeclareUnicodeCharacter{8759}{::}
\DeclareUnicodeCharacter{9669}{\ensuremath{\triangleleft}}
\DeclareUnicodeCharacter{8799}{\ensuremath{\stackrel{\scriptscriptstyle ?}{=}}}
\DeclareUnicodeCharacter{10214}{\ensuremath{\llbracket}}
\DeclareUnicodeCharacter{10215}{\ensuremath{\rrbracket}}

\renewcommand\Varid[1]{\mathord{\textsf{#1}}}
\let\Conid\Varid
\newcommand\Keyword[1]{\textsf{\textbf{#1}}}
\EndFmtInput

\usepackage[usenames,dvipsnames]{color}
\usepackage[
  colorlinks = true,
  linkcolor  = Maroon,
  urlcolor   = MidnightBlue,
  citecolor  = ForestGreen
  ]{hyperref}

\usepackage[cm]{fullpage}
\usepackage{natbib}
\makeatletter
\renewcommand\bibsection%
{
  \section*{\refname
    \@mkboth{\MakeUppercase{\refname}}{\MakeUppercase{\refname}}}
}
\makeatother

\usepackage{ifthen}
\usepackage{enumitem}

\newboolean{showNotes}
\setboolean{showNotes}{true}

\newcommand{\todo}[1]{
  \ifthenelse%
  {\boolean{showNotes}}
  {\textcolor{red}{\textbf{TODO:~}#1}}
  {}}

\usepackage{etoolbox}%
\newboolean{partial}
\setboolean{partial}{false}
\AtEndEnvironment{hscode}{\ifthenelse{\boolean{partial}}{\\...}{}}

\newboolean{showHidden}
\setboolean{showHidden}{false}

\usepackage{mdframed}
\usepackage{bussproofs}
\EnableBpAbbreviations%
\def\fCenter{\ \vdash\ }

\newenvironment{scprooftree}[1]%
  {\gdef\scalefactor{#1}\begin{center}\proofSkipAmount \leavevmode}%
  {\scalebox{\scalefactor}{\DisplayProof}\proofSkipAmount \end{center} }

\usepackage{textgreek}
\usepackage{stmaryrd}
\usepackage{scalerel}
\usepackage{pict2e}
\usepackage{picture}
\usepackage{ucs}
\usepackage[utf8x]{inputenc}

\newcommand{\varslash}{%
  \mathbin{\mathpalette\pictslash{{0}{1}}}%
}
\newcommand{\varbslash}{%
  \mathbin{\mathpalette\pictslash{{1}{-1}}}%
}

\usepackage{etoolbox}
\patchcmd{\thebibliography}{\chapter*}{\section*}{}{}

\makeatletter
\newcommand{\pictslash}[2]{%
  \vcenter{%
    \sbox0{$\m@th#1\varobslash$}\dimen0=.55\wd0
    \hbox to\wd 0{%
      \hfil\pictslash@aux#2\hfil
    }%
  }%
}
\newcommand{\pictslash@aux}[2]{%
    \begin{picture}(\dimen0,\dimen0)
    \roundcap
    \linethickness{.15ex}
    \put(0,#1\dimen0){\line(1,#2){\dimen0}}
    \end{picture}%
}
\makeatother

\newcommand{\holer}[1]{#1{\scriptscriptstyle\triangleright}}
\newcommand{\holel}[1]{{\scriptscriptstyle\triangleleft}#1}

\DeclareUnicodeCharacter{8656}{$\varslash$}
\DeclareUnicodeCharacter{8658}{$\varbslash$}
\DeclareUnicodeCharacter{8666}{$\varoslash$}
\DeclareUnicodeCharacter{8667}{$\varobslash$}
\DeclareUnicodeCharacter{7478}{$^{J}$}
\DeclareUnicodeCharacter{7480}{$^{L}$}
\DeclareUnicodeCharacter{7487}{$^{R}$}
\DeclareUnicodeCharacter{7489}{$^{U}$}
\DeclareUnicodeCharacter{7522}{$_{i}$}
\DeclareUnicodeCharacter{8314}{$^{+}$}
\DeclareUnicodeCharacter{8315}{$^{-}$}
\DeclareUnicodeCharacter{9633}{$\Box$}
\DeclareUnicodeCharacter{9671}{$\Diamond$}

\renewcommand{\hscodestyle}{%
  \setlength\leftskip{\parindent}%
  \footnotesize
}

\begin{document}

\title{Formalising type-logical grammars in Agda}%
\author{Wen Kokke}%
\date{March, 2015}
\institute{Utrecht University}%
\maketitle

\begin{abstract}
In recent years, the interest in using proof assistants to formalise
and reason about mathematics and programming languages has grown.
Type-logical grammars, being closely related to type theories and
systems used in functional programming, are a perfect candidate to
next apply this curiosity to.
The advantages of using proof assistants is that they allow one to
write formally verified proofs about one's type-logical systems, and
that any theory, once implemented, can immediately be computed with.
The downside is that in many cases the formal proofs are written as an
afterthought, are incomplete, or use obtuse syntax.
This makes it that the verified proofs are often much more difficult
to read than the pen-and-paper proofs, and almost never directly
published.
In this paper, we will try to remedy that by example.

Concretely, we use Agda to model the Lambek-Grishin calculus, a
grammar logic with a rich vocabulary of type-forming operations.
We then present a verified procedure for cut elimination in this
system. Then we briefly outline a CPS translation from proofs in the
Lambek-Grishin calculus to programs in Agda. And finally, we will put
our system to use in the analysis of a simple example sentence.
\end{abstract}

\section{Introduction}

Why would we want to formalise type-logical grammars using proof
assistants? One good reason is that it allows us to write formally
verified proofs about the theoretical properties of our type-logical
grammars. But not only that---it allows us to directly run our proofs
as programs. For instance, we can directly run the procedure for cut
elimination in this paper to investigate what kind of derivations are
created by it \textit{and} be confident in its correctness.

Why, then, would we want to use Agda instead of a more established
proof assistant such as, for instance, Coq? There are several good
reasons, but we believe that the syntactic freedom offered by Agda is
the most important.
It is this freedom that allows us to write machine-checkable proofs,
formatted in a way which is very close to the way one would otherwise
typeset proofs, and which are highly readable compared to other
machine-checked proofs.
This means that we can be confident that the proofs \textit{as they
are published} are correct, and that they are necessarily complete---for
though we can hide some of the less interesting definitions from the
final paper, we cannot omit them from the source.

Additionally, because there is a one-to-one correspondence between the
published proofs and the code, it becomes very easy for the reader to
start up a proof environment and inspect the proofs interactively in
order to further their understanding.

Our test case in this paper is the Lambek-Grishin calculus \citep[LG,
][]{moortgat2009}. LG is an example of an extended Lambek calculus. In
addition to the product (\ensuremath{\Varid{⊗}}) and the residual slashes (\ensuremath{\Varid{⇒}}, \ensuremath{\Varid{⇐}}), LG
has a dual family with \ensuremath{\Varid{⊕}} and difference operations (\ensuremath{\Varid{⇚}}, \ensuremath{\Varid{⇛}})
together with distributivity principles for the interaction between
the two families. See \citet{moortgat2009} for discussion of how LG
overcomes syntactic and semantic limitations of the original Lambek
calculus.

We will formalise the residuation-monotonicity axiomatisation for the
Lambek-Grishin calculus~\citep{moortgat2007} in Agda, present a
verified procedure for cut elimination in this system, and briefly
outline a CPS translation into Agda. There are several reasons why we
have chosen to formalise this particular system.
\begin{itemize}
\item%
  It allows cut as an admissible rule, i.e.\ a function on proofs,
  instead of defining a separate cut-free system and a cut-elimination
  procedure;
\item%
  it has efficiently decidable proof search, largely due to the
  absence of the cut rule;
\item%
  it has some interesting symmetries, as explored in
  \citet{moortgat2007,moortgat2009}. Because of this, most proofs of
  properties of LG are not much more complicated than their associated
  proofs for the non-associative Lambek calculus;
\item
  it has a continuation-passing style interpretation, which has shown to
  be useful in both derivational and lexical semantics
  (\citeauthor{moortgat2007,bs2015,asher2011});
\item%
  lastly, an implementation of the non-associative Lambek calculus can
  easily and mechanically be extracted from our implementation of LG.
\end{itemize}
Since this paper is by no means a complete introduction to Agda or to
dependently-typed programming, we advise the interested reader to
refer to \citet{norell2009} for a detailed discussion of Agda.

It should be mentioned that (although we omit some of the more tedious
parts) this paper is written in literate Agda, and the code has been
made available on GitHub.\footnote{\href%
{https://gist.github.com/wenkokke/cc12b92a8a60696b712c\#file-main-agda}%
{gist.github.com/wenkokke/cc12b92a8a60696b712c}}

\section{Formulas, Judgements, Base System}

If we want to model our type-logical grammars in Agda, a natural
starting point would be our atomic formulas---such as \ensuremath{\Varid{n}}, \ensuremath{\Varid{np}}, \ensuremath{\Varid{s}},
etc. These could easily be represented as an enumerated data
type. However, in order to avoid committing to a certain set of atomic
formulas and side-step the debate on which formulas \textit{should} be
atomic, we will simply assume there is a some data type representing
our atomic formulas. This will be reflected in our module header as
follows:
\begin{hscode}\SaveRestoreHook
\column{B}{@{}>{\hspre}l<{\hspost}@{}}%
\column{E}{@{}>{\hspre}l<{\hspost}@{}}%
\>[B]{}\Keyword{module}\;\Varid{logic}\;(\Conid{Univ}\;\mathbin{:}\;\Conid{Set})\;\Keyword{where}{}\<[E]%
\ColumnHook
\end{hscode}\resethooks
Our formulas can easily be described as a data type, injecting our
atomic formulas by means of the constructor \ensuremath{\Varid{el}}, and adding the
familiar connectives from the Lambek-Grishin calculus as binary
constructors. Note that, in Agda, we can use underscores in
definitions to denote argument positions. This means that \ensuremath{\Varid{\char95 ⊗\char95 }} below
defines an infix, binary connective:
\begin{hscode}\SaveRestoreHook
\column{B}{@{}>{\hspre}l<{\hspost}@{}}%
\column{3}{@{}>{\hspre}l<{\hspost}@{}}%
\column{5}{@{}>{\hspre}l<{\hspost}@{}}%
\column{18}{@{}>{\hspre}l<{\hspost}@{}}%
\column{26}{@{}>{\hspre}l<{\hspost}@{}}%
\column{34}{@{}>{\hspre}l<{\hspost}@{}}%
\column{E}{@{}>{\hspre}l<{\hspost}@{}}%
\>[3]{}\Keyword{data}\;\Conid{Type}\;\mathbin{:}\;\Conid{Set}\;\Keyword{where}{}\<[E]%
\\
\>[3]{}\hsindent{2}{}\<[5]%
\>[5]{}\Varid{el}\;{}\<[18]%
\>[18]{}\mathbin{:}\;\Conid{Univ}\;{}\<[26]%
\>[26]{}\Varid{→}\;\Conid{Type}{}\<[E]%
\\
\>[3]{}\hsindent{2}{}\<[5]%
\>[5]{}\Varid{\char95 ⊗\char95 }\;\Varid{\char95 ⇒\char95 }\;\Varid{\char95 ⇐\char95 }\;{}\<[18]%
\>[18]{}\mathbin{:}\;\Conid{Type}\;{}\<[26]%
\>[26]{}\Varid{→}\;\Conid{Type}\;{}\<[34]%
\>[34]{}\Varid{→}\;\Conid{Type}{}\<[E]%
\\
\>[3]{}\hsindent{2}{}\<[5]%
\>[5]{}\Varid{\char95 ⊕\char95 }\;\Varid{\char95 ⇚\char95 }\;\Varid{\char95 ⇛\char95 }\;{}\<[18]%
\>[18]{}\mathbin{:}\;\Conid{Type}\;{}\<[26]%
\>[26]{}\Varid{→}\;\Conid{Type}\;{}\<[34]%
\>[34]{}\Varid{→}\;\Conid{Type}{}\<[E]%
\ColumnHook
\end{hscode}\resethooks
In the same manner, we can define a data type to represent judgements:
\begin{hscode}\SaveRestoreHook
\column{B}{@{}>{\hspre}l<{\hspost}@{}}%
\column{3}{@{}>{\hspre}l<{\hspost}@{}}%
\column{5}{@{}>{\hspre}l<{\hspost}@{}}%
\column{10}{@{}>{\hspre}l<{\hspost}@{}}%
\column{E}{@{}>{\hspre}l<{\hspost}@{}}%
\>[3]{}\Keyword{data}\;\Conid{Judgement}\;\mathbin{:}\;\Conid{Set}\;\Keyword{where}{}\<[E]%
\\
\>[3]{}\hsindent{2}{}\<[5]%
\>[5]{}\Varid{\char95 ⊢\char95 }\;{}\<[10]%
\>[10]{}\mathbin{:}\;\Conid{Type}\;\Varid{→}\;\Conid{Type}\;\Varid{→}\;\Conid{Judgement}{}\<[E]%
\ColumnHook
\end{hscode}\resethooks
Using the above definitions, we can now write judgements such as \ensuremath{\Conid{A}\;\Varid{⊗}\;\Conid{A}\;\Varid{⇒}\;\Conid{B}\;\Varid{⊢}\;\Conid{B}} as Agda values.
Next we will define a data type to represent our logical system. This
is where we can use the dependent type system to our advantage. The
constructors for our data type will represent the axiomatic inference
rules of the system, and their \textit{types} will be constrained by
judgements. Below you can see the entire system LG as an Agda data
type\footnote{%
  For the typeset version of this paper we omit the quantifiers for
  all implicit, universally quantified arguments.
}:
\begin{hscode}\SaveRestoreHook
\column{B}{@{}>{\hspre}l<{\hspost}@{}}%
\column{3}{@{}>{\hspre}l<{\hspost}@{}}%
\column{5}{@{}>{\hspre}l<{\hspost}@{}}%
\column{10}{@{}>{\hspre}l<{\hspost}@{}}%
\column{13}{@{}>{\hspre}l<{\hspost}@{}}%
\column{23}{@{}>{\hspre}l<{\hspost}@{}}%
\column{27}{@{}>{\hspre}l<{\hspost}@{}}%
\column{31}{@{}>{\hspre}l<{\hspost}@{}}%
\column{35}{@{}>{\hspre}l<{\hspost}@{}}%
\column{E}{@{}>{\hspre}l<{\hspost}@{}}%
\>[3]{}\Keyword{data}\;\mathit{LG}\_\;\mathbin{:}\;\Conid{Judgement}\;\Varid{→}\;\Conid{Set}\;\Keyword{where}{}\<[E]%
\\[\blanklineskip]%
\>[3]{}\hsindent{2}{}\<[5]%
\>[5]{}\Varid{ax}\;{}\<[10]%
\>[10]{}\mathbin{:}\;{}\<[13]%
\>[13]{}\mathit{LG}\;\Varid{el}\;\Conid{A}\;\Varid{⊢}\;\Varid{el}\;\Conid{A}{}\<[E]%
\\[\blanklineskip]%
\>[3]{}\hsindent{2}{}\<[5]%
\>[5]{}\mbox{\onelinecomment  residuation and monotonicity for (⇐ , ⊗ , ⇒)}{}\<[E]%
\\
\>[3]{}\hsindent{2}{}\<[5]%
\>[5]{}\Varid{r⇒⊗}\;{}\<[10]%
\>[10]{}\mathbin{:}\;{}\<[13]%
\>[13]{}\mathit{LG}\;\Conid{B}\;\Varid{⊢}\;\Conid{A}\;\Varid{⇒}\;\Conid{C}\;{}\<[27]%
\>[27]{}\Varid{→}\;\mathit{LG}\;\Conid{A}\;\Varid{⊗}\;\Conid{B}\;\Varid{⊢}\;\Conid{C}{}\<[E]%
\\
\>[3]{}\hsindent{2}{}\<[5]%
\>[5]{}\Varid{r⊗⇒}\;{}\<[10]%
\>[10]{}\mathbin{:}\;{}\<[13]%
\>[13]{}\mathit{LG}\;\Conid{A}\;\Varid{⊗}\;\Conid{B}\;\Varid{⊢}\;\Conid{C}\;{}\<[27]%
\>[27]{}\Varid{→}\;\mathit{LG}\;\Conid{B}\;\Varid{⊢}\;\Conid{A}\;\Varid{⇒}\;\Conid{C}{}\<[E]%
\\
\>[3]{}\hsindent{2}{}\<[5]%
\>[5]{}\Varid{r⇐⊗}\;{}\<[10]%
\>[10]{}\mathbin{:}\;{}\<[13]%
\>[13]{}\mathit{LG}\;\Conid{A}\;\Varid{⊢}\;\Conid{C}\;\Varid{⇐}\;\Conid{B}\;{}\<[27]%
\>[27]{}\Varid{→}\;\mathit{LG}\;\Conid{A}\;\Varid{⊗}\;\Conid{B}\;\Varid{⊢}\;\Conid{C}{}\<[E]%
\\
\>[3]{}\hsindent{2}{}\<[5]%
\>[5]{}\Varid{r⊗⇐}\;{}\<[10]%
\>[10]{}\mathbin{:}\;{}\<[13]%
\>[13]{}\mathit{LG}\;\Conid{A}\;\Varid{⊗}\;\Conid{B}\;\Varid{⊢}\;\Conid{C}\;{}\<[27]%
\>[27]{}\Varid{→}\;\mathit{LG}\;\Conid{A}\;\Varid{⊢}\;\Conid{C}\;\Varid{⇐}\;\Conid{B}{}\<[E]%
\\[\blanklineskip]%
\>[3]{}\hsindent{2}{}\<[5]%
\>[5]{}\Varid{m⊗}\;{}\<[10]%
\>[10]{}\mathbin{:}\;{}\<[13]%
\>[13]{}\mathit{LG}\;\Conid{A}\;\Varid{⊢}\;\Conid{B}\;{}\<[23]%
\>[23]{}\Varid{→}\;\mathit{LG}\;\Conid{C}\;\Varid{⊢}\;\Conid{D}\;{}\<[35]%
\>[35]{}\Varid{→}\;\mathit{LG}\;\Conid{A}\;\Varid{⊗}\;\Conid{C}\;\Varid{⊢}\;\Conid{B}\;\Varid{⊗}\;\Conid{D}{}\<[E]%
\\
\>[3]{}\hsindent{2}{}\<[5]%
\>[5]{}\Varid{m⇒}\;{}\<[10]%
\>[10]{}\mathbin{:}\;{}\<[13]%
\>[13]{}\mathit{LG}\;\Conid{A}\;\Varid{⊢}\;\Conid{B}\;{}\<[23]%
\>[23]{}\Varid{→}\;\mathit{LG}\;\Conid{C}\;\Varid{⊢}\;\Conid{D}\;{}\<[35]%
\>[35]{}\Varid{→}\;\mathit{LG}\;\Conid{B}\;\Varid{⇒}\;\Conid{C}\;\Varid{⊢}\;\Conid{A}\;\Varid{⇒}\;\Conid{D}{}\<[E]%
\\
\>[3]{}\hsindent{2}{}\<[5]%
\>[5]{}\Varid{m⇐}\;{}\<[10]%
\>[10]{}\mathbin{:}\;{}\<[13]%
\>[13]{}\mathit{LG}\;\Conid{A}\;\Varid{⊢}\;\Conid{B}\;{}\<[23]%
\>[23]{}\Varid{→}\;\mathit{LG}\;\Conid{C}\;\Varid{⊢}\;\Conid{D}\;{}\<[35]%
\>[35]{}\Varid{→}\;\mathit{LG}\;\Conid{A}\;\Varid{⇐}\;\Conid{D}\;\Varid{⊢}\;\Conid{B}\;\Varid{⇐}\;\Conid{C}{}\<[E]%
\\[\blanklineskip]%
\>[3]{}\hsindent{2}{}\<[5]%
\>[5]{}\mbox{\onelinecomment  residuation and monotonicity for (⇚ , ⊕ , ⇛)}{}\<[E]%
\\
\>[3]{}\hsindent{2}{}\<[5]%
\>[5]{}\Varid{r⇛⊕}\;{}\<[10]%
\>[10]{}\mathbin{:}\;{}\<[13]%
\>[13]{}\mathit{LG}\;\Conid{B}\;\Varid{⇛}\;\Conid{C}\;\Varid{⊢}\;\Conid{A}\;{}\<[27]%
\>[27]{}\Varid{→}\;\mathit{LG}\;\Conid{C}\;\Varid{⊢}\;\Conid{B}\;\Varid{⊕}\;\Conid{A}{}\<[E]%
\\
\>[3]{}\hsindent{2}{}\<[5]%
\>[5]{}\Varid{r⊕⇛}\;{}\<[10]%
\>[10]{}\mathbin{:}\;{}\<[13]%
\>[13]{}\mathit{LG}\;\Conid{C}\;\Varid{⊢}\;\Conid{B}\;\Varid{⊕}\;\Conid{A}\;{}\<[27]%
\>[27]{}\Varid{→}\;\mathit{LG}\;\Conid{B}\;\Varid{⇛}\;\Conid{C}\;\Varid{⊢}\;\Conid{A}{}\<[E]%
\\
\>[3]{}\hsindent{2}{}\<[5]%
\>[5]{}\Varid{r⊕⇚}\;{}\<[10]%
\>[10]{}\mathbin{:}\;{}\<[13]%
\>[13]{}\mathit{LG}\;\Conid{C}\;\Varid{⊢}\;\Conid{B}\;\Varid{⊕}\;\Conid{A}\;{}\<[27]%
\>[27]{}\Varid{→}\;\mathit{LG}\;\Conid{C}\;\Varid{⇚}\;\Conid{A}\;\Varid{⊢}\;\Conid{B}{}\<[E]%
\\
\>[3]{}\hsindent{2}{}\<[5]%
\>[5]{}\Varid{r⇚⊕}\;{}\<[10]%
\>[10]{}\mathbin{:}\;{}\<[13]%
\>[13]{}\mathit{LG}\;\Conid{C}\;\Varid{⇚}\;\Conid{A}\;\Varid{⊢}\;\Conid{B}\;{}\<[27]%
\>[27]{}\Varid{→}\;\mathit{LG}\;\Conid{C}\;\Varid{⊢}\;\Conid{B}\;\Varid{⊕}\;\Conid{A}{}\<[E]%
\\[\blanklineskip]%
\>[3]{}\hsindent{2}{}\<[5]%
\>[5]{}\Varid{m⊕}\;{}\<[10]%
\>[10]{}\mathbin{:}\;{}\<[13]%
\>[13]{}\mathit{LG}\;\Conid{A}\;\Varid{⊢}\;\Conid{B}\;{}\<[23]%
\>[23]{}\Varid{→}\;\mathit{LG}\;\Conid{C}\;\Varid{⊢}\;\Conid{D}\;{}\<[35]%
\>[35]{}\Varid{→}\;\mathit{LG}\;\Conid{A}\;\Varid{⊕}\;\Conid{C}\;\Varid{⊢}\;\Conid{B}\;\Varid{⊕}\;\Conid{D}{}\<[E]%
\\
\>[3]{}\hsindent{2}{}\<[5]%
\>[5]{}\Varid{m⇛}\;{}\<[10]%
\>[10]{}\mathbin{:}\;{}\<[13]%
\>[13]{}\mathit{LG}\;\Conid{C}\;\Varid{⊢}\;\Conid{D}\;{}\<[23]%
\>[23]{}\Varid{→}\;\mathit{LG}\;\Conid{A}\;\Varid{⊢}\;\Conid{B}\;{}\<[35]%
\>[35]{}\Varid{→}\;\mathit{LG}\;\Conid{D}\;\Varid{⇛}\;\Conid{A}\;\Varid{⊢}\;\Conid{C}\;\Varid{⇛}\;\Conid{B}{}\<[E]%
\\
\>[3]{}\hsindent{2}{}\<[5]%
\>[5]{}\Varid{m⇚}\;{}\<[10]%
\>[10]{}\mathbin{:}\;{}\<[13]%
\>[13]{}\mathit{LG}\;\Conid{A}\;\Varid{⊢}\;\Conid{B}\;{}\<[23]%
\>[23]{}\Varid{→}\;\mathit{LG}\;\Conid{C}\;\Varid{⊢}\;\Conid{D}\;{}\<[35]%
\>[35]{}\Varid{→}\;\mathit{LG}\;\Conid{A}\;\Varid{⇚}\;\Conid{D}\;\Varid{⊢}\;\Conid{B}\;\Varid{⇚}\;\Conid{C}{}\<[E]%
\\[\blanklineskip]%
\>[3]{}\hsindent{2}{}\<[5]%
\>[5]{}\mbox{\onelinecomment  grishin distributives}{}\<[E]%
\\
\>[3]{}\hsindent{2}{}\<[5]%
\>[5]{}\Varid{d⇛⇐}\;{}\<[10]%
\>[10]{}\mathbin{:}\;{}\<[13]%
\>[13]{}\mathit{LG}\;\Conid{A}\;\Varid{⊗}\;\Conid{B}\;\Varid{⊢}\;\Conid{C}\;\Varid{⊕}\;\Conid{D}\;{}\<[31]%
\>[31]{}\Varid{→}\;\mathit{LG}\;\Conid{C}\;\Varid{⇛}\;\Conid{A}\;\Varid{⊢}\;\Conid{D}\;\Varid{⇐}\;\Conid{B}{}\<[E]%
\\
\>[3]{}\hsindent{2}{}\<[5]%
\>[5]{}\Varid{d⇛⇒}\;{}\<[10]%
\>[10]{}\mathbin{:}\;{}\<[13]%
\>[13]{}\mathit{LG}\;\Conid{A}\;\Varid{⊗}\;\Conid{B}\;\Varid{⊢}\;\Conid{C}\;\Varid{⊕}\;\Conid{D}\;{}\<[31]%
\>[31]{}\Varid{→}\;\mathit{LG}\;\Conid{C}\;\Varid{⇛}\;\Conid{B}\;\Varid{⊢}\;\Conid{A}\;\Varid{⇒}\;\Conid{D}{}\<[E]%
\\
\>[3]{}\hsindent{2}{}\<[5]%
\>[5]{}\Varid{d⇚⇒}\;{}\<[10]%
\>[10]{}\mathbin{:}\;{}\<[13]%
\>[13]{}\mathit{LG}\;\Conid{A}\;\Varid{⊗}\;\Conid{B}\;\Varid{⊢}\;\Conid{C}\;\Varid{⊕}\;\Conid{D}\;{}\<[31]%
\>[31]{}\Varid{→}\;\mathit{LG}\;\Conid{B}\;\Varid{⇚}\;\Conid{D}\;\Varid{⊢}\;\Conid{A}\;\Varid{⇒}\;\Conid{C}{}\<[E]%
\\
\>[3]{}\hsindent{2}{}\<[5]%
\>[5]{}\Varid{d⇚⇐}\;{}\<[10]%
\>[10]{}\mathbin{:}\;{}\<[13]%
\>[13]{}\mathit{LG}\;\Conid{A}\;\Varid{⊗}\;\Conid{B}\;\Varid{⊢}\;\Conid{C}\;\Varid{⊕}\;\Conid{D}\;{}\<[31]%
\>[31]{}\Varid{→}\;\mathit{LG}\;\Conid{A}\;\Varid{⇚}\;\Conid{D}\;\Varid{⊢}\;\Conid{C}\;\Varid{⇐}\;\Conid{B}{}\<[E]%
\ColumnHook
\end{hscode}\resethooks
Note that Agda allows arbitrary unicode characters in identifiers, so
\ensuremath{\Varid{r⊗⇒}} is a valid Agda identifier.

Using this data type, we can already do quite a lot. For instance, we
can show that while the inference rule \ensuremath{\Varid{ax}} above is restricted to
atomic formulas\footnote{%
  Whereas the rule \ensuremath{\Varid{ax}} may appear to be unrestricted, it only allows
  the derivation of the identity proof for any formula \ensuremath{\Varid{el}\;\Conid{A}}. That
  is, any \textit{atomic formula} \ensuremath{\Conid{A}} delimited by the constructor
  \ensuremath{\Varid{el}}.
}, the unrestricted version is admissible, by induction on
the formula. Note that the construction \ensuremath{\{\mskip1.5mu \Conid{A}\;\mathrel{=}\;\Varid{...}\mskip1.5mu\}} below is used to
pattern match on the implicit variable \ensuremath{\Conid{A}}:
\begin{hscode}\SaveRestoreHook
\column{B}{@{}>{\hspre}l<{\hspost}@{}}%
\column{3}{@{}>{\hspre}l<{\hspost}@{}}%
\column{17}{@{}>{\hspre}l<{\hspost}@{}}%
\column{E}{@{}>{\hspre}l<{\hspost}@{}}%
\>[3]{}\Varid{ax′}\;\mathbin{:}\;\mathit{LG}\;\Conid{A}\;\Varid{⊢}\;\Conid{A}{}\<[E]%
\\
\>[3]{}\Varid{ax′}\;\{\mskip1.5mu \Conid{A}\;\mathrel{=}\;\Varid{el}\;{}\<[17]%
\>[17]{}\anonymous \mskip1.5mu\}\;\mathrel{=}\;\Varid{ax}{}\<[E]%
\\
\>[3]{}\Varid{ax′}\;\{\mskip1.5mu \Conid{A}\;\mathrel{=}\;\anonymous \;\Varid{⊗}\;{}\<[17]%
\>[17]{}\anonymous \mskip1.5mu\}\;\mathrel{=}\;\Varid{m⊗}\;\Varid{ax′}\;\Varid{ax′}{}\<[E]%
\\
\>[3]{}\Varid{ax′}\;\{\mskip1.5mu \Conid{A}\;\mathrel{=}\;\anonymous \;\Varid{⇐}\;{}\<[17]%
\>[17]{}\anonymous \mskip1.5mu\}\;\mathrel{=}\;\Varid{m⇐}\;\Varid{ax′}\;\Varid{ax′}{}\<[E]%
\\
\>[3]{}\Varid{ax′}\;\{\mskip1.5mu \Conid{A}\;\mathrel{=}\;\anonymous \;\Varid{⇒}\;{}\<[17]%
\>[17]{}\anonymous \mskip1.5mu\}\;\mathrel{=}\;\Varid{m⇒}\;\Varid{ax′}\;\Varid{ax′}{}\<[E]%
\\
\>[3]{}\Varid{ax′}\;\{\mskip1.5mu \Conid{A}\;\mathrel{=}\;\anonymous \;\Varid{⊕}\;{}\<[17]%
\>[17]{}\anonymous \mskip1.5mu\}\;\mathrel{=}\;\Varid{m⊕}\;\Varid{ax′}\;\Varid{ax′}{}\<[E]%
\\
\>[3]{}\Varid{ax′}\;\{\mskip1.5mu \Conid{A}\;\mathrel{=}\;\anonymous \;\Varid{⇚}\;{}\<[17]%
\>[17]{}\anonymous \mskip1.5mu\}\;\mathrel{=}\;\Varid{m⇚}\;\Varid{ax′}\;\Varid{ax′}{}\<[E]%
\\
\>[3]{}\Varid{ax′}\;\{\mskip1.5mu \Conid{A}\;\mathrel{=}\;\anonymous \;\Varid{⇛}\;{}\<[17]%
\>[17]{}\anonymous \mskip1.5mu\}\;\mathrel{=}\;\Varid{m⇛}\;\Varid{ax′}\;\Varid{ax′}{}\<[E]%
\ColumnHook
\end{hscode}\resethooks
Alternatively, we could derive the various applications and
co-applications that hold in the Lambek-Grishin calculus:
\begin{hscode}\SaveRestoreHook
\column{B}{@{}>{\hspre}l<{\hspost}@{}}%
\column{3}{@{}>{\hspre}l<{\hspost}@{}}%
\column{12}{@{}>{\hspre}l<{\hspost}@{}}%
\column{E}{@{}>{\hspre}l<{\hspost}@{}}%
\>[3]{}\Varid{appl-⇒′}\;{}\<[12]%
\>[12]{}\mathbin{:}\;\mathit{LG}\;\Conid{A}\;\Varid{⊗}\;(\Conid{A}\;\Varid{⇒}\;\Conid{B})\;\Varid{⊢}\;\Conid{B}{}\<[E]%
\\
\>[3]{}\Varid{appl-⇒′}\;{}\<[12]%
\>[12]{}\mathrel{=}\;\Varid{r⇒⊗}\;(\Varid{m⇒}\;\Varid{ax′}\;\Varid{ax′}){}\<[E]%
\\[\blanklineskip]%
\>[3]{}\Varid{appl-⇐′}\;{}\<[12]%
\>[12]{}\mathbin{:}\;\mathit{LG}\;(\Conid{B}\;\Varid{⇐}\;\Conid{A})\;\Varid{⊗}\;\Conid{A}\;\Varid{⊢}\;\Conid{B}{}\<[E]%
\\
\>[3]{}\Varid{appl-⇐′}\;{}\<[12]%
\>[12]{}\mathrel{=}\;\Varid{r⇐⊗}\;(\Varid{m⇐}\;\Varid{ax′}\;\Varid{ax′}){}\<[E]%
\\[\blanklineskip]%
\>[3]{}\Varid{appl-⇛′}\;{}\<[12]%
\>[12]{}\mathbin{:}\;\mathit{LG}\;\Conid{B}\;\Varid{⊢}\;\Conid{A}\;\Varid{⊕}\;(\Conid{A}\;\Varid{⇛}\;\Conid{B}){}\<[E]%
\\
\>[3]{}\Varid{appl-⇛′}\;{}\<[12]%
\>[12]{}\mathrel{=}\;\Varid{r⇛⊕}\;(\Varid{m⇛}\;\Varid{ax′}\;\Varid{ax′}){}\<[E]%
\\[\blanklineskip]%
\>[3]{}\Varid{appl-⇚′}\;{}\<[12]%
\>[12]{}\mathbin{:}\;\mathit{LG}\;\Conid{B}\;\Varid{⊢}\;(\Conid{B}\;\Varid{⇚}\;\Conid{A})\;\Varid{⊕}\;\Conid{A}{}\<[E]%
\\
\>[3]{}\Varid{appl-⇚′}\;{}\<[12]%
\>[12]{}\mathrel{=}\;\Varid{r⇚⊕}\;(\Varid{m⇚}\;\Varid{ax′}\;\Varid{ax′}){}\<[E]%
\ColumnHook
\end{hscode}\resethooks
However, the most compelling reason to use the axiomatisation we have
chosen, using residuation and monotonicity rules, is that cut becomes
an admissible rule.

\section{Admissible Cut}

We would like to show that \ensuremath{\Varid{cut′}} of type \ensuremath{\mathit{LG}\;\Conid{A}\;\Varid{⊢}\;\Conid{B}\;\Varid{→}\;\mathit{LG}\;\Conid{B}\;\Varid{⊢}\;\Conid{C}\;\Varid{→}\;\mathit{LG}\;\Conid{A}\;\Varid{⊢}\;\Conid{C}} is
an admissible rule.
The method of \citet{moortgat1999}, for the basic non-associative
Lambek calculus, can be readily generalized to the case of LG:
\begin{enumerate}[label= (\roman*)]
\item\label{p1} every connective is introduced \textit{symmetrically} by a
  monotonicity rule or as an axiom;
\item\label{p2} every connective has one side (antecedent or succedent) where,
  if it occurs there at the top level, it cannot be taken apart or
  changed by any inference rule;
\item\label{p3} as a consequence of~\ref{p2}, every formula has one
  side where, if it occurs there at the top level, it is immutable,
  i.e.\ there is no rule   which can eliminate it;
\item\label{p4} due to~\ref{p1} and~\ref{p3}, when we find such an
  immutable formula, we can be sure that, stepping through the
  derivation, after some number of steps we will find the monotonicity
  rule which introduced that formula;
\item\label{p5} due to the type of \ensuremath{\Varid{cut′}}, when we match on the cut
  formula \ensuremath{\Conid{B}} we will always have an immutable variant of that formula
  in either the first or the second argument of \ensuremath{\Varid{cut′}};
\item\label{p6} finally, for each main connective there exists a
  rewrite rule which makes use of the facts in~\ref{p4} and~\ref{p5}
  to rewrite an application of \ensuremath{\Varid{cut′}}: to two applications of \ensuremath{\Varid{cut′}}
  on the arguments of the monotonicity rule which introduced the
  connective, chained together by applications of residuation (for
  binary connectives) or simply to a derivation (for atomic
  formulas).  As an example, the rewrite rule for \ensuremath{\Varid{\char95 ⊗\char95 }} can be found
  in figure~\ref{cut:otimes}.
\end{enumerate}
\begin{figure*}[ht]%
  \footnotesize
  \hspace*{ -\parindent }%
  \begin{minipage}{.47\linewidth}
    \begin{prooftree}
      \AXC{$E     \vdash B    $}
      \AXC{$    F \vdash     C$}
      \BIC{$E ⊗ F \vdash B ⊗ C$}
      \UIC{$      \vdots      $}
      \UIC{$A     \vdash B ⊗ C$}
      \AXC{$B ⊗ C \vdash D    $}
      \BIC{$A     \vdash D    $}
    \end{prooftree}
  \end{minipage}
  \begin{minipage}{.06\linewidth}
    $$\Longrightarrow$$
  \end{minipage}
  \begin{minipage}{.47\linewidth}
    \begin{prooftree}
      \footnotesize
      \AXC{$E     \vdash B    $}
      \AXC{$    F \vdash C    $}
      \AXC{$B ⊗ C \vdash     D$}
      \UIC{$    C \vdash B \varbslash D$}
      \BIC{$    F \vdash B \varbslash D$}
      \UIC{$B ⊗ F \vdash     D$}
      \UIC{$B     \vdash D \varslash F$}
      \BIC{$E     \vdash D \varslash F$}
      \UIC{$E ⊗ F \vdash D    $}
      \UIC{$      \vdots      $}
      \UIC{$A     \vdash D    $}
    \end{prooftree}
  \end{minipage}%
\caption{Rewrite rule for cut on formula \ensuremath{\Conid{B}\;\Varid{⊗}\;\Conid{C}}.}
\label{cut:otimes}
\end{figure*}
We can model the view on the left-hand side of the rewrite rule in
figure~\ref{cut:otimes} as a data type.
In order to construct this view for some suitable derivation \ensuremath{\Varid{f}}, we
need two derivations, \ensuremath{\Varid{h₁}} and \ensuremath{\Varid{h₂}} and a derivation \ensuremath{\Varid{f′}}, which
represents the arbitrary derivation steps taking \ensuremath{(\Varid{m⊗}\;\Varid{h₁}\;\Varid{h₂})} back to
\ensuremath{\Varid{f}}. Lastly, we include a proof \ensuremath{\Varid{pr}} of the fact that the
reconstructed derivation \ensuremath{\Varid{f′}\;(\Varid{m⊗}\;\Varid{h₁}\;\Varid{h₂})} is identical to \ensuremath{\Varid{f}}:
\begin{hscode}\SaveRestoreHook
\column{B}{@{}>{\hspre}l<{\hspost}@{}}%
\column{3}{@{}>{\hspre}l<{\hspost}@{}}%
\column{5}{@{}>{\hspre}l<{\hspost}@{}}%
\column{15}{@{}>{\hspre}l<{\hspost}@{}}%
\column{20}{@{}>{\hspre}l<{\hspost}@{}}%
\column{E}{@{}>{\hspre}l<{\hspost}@{}}%
\>[3]{}\Keyword{data}\;\Conid{Origin}\;(\Varid{f}\;{}\<[20]%
\>[20]{}\mathbin{:}\;\mathit{LG}\;\Conid{A}\;\Varid{⊢}\;\Conid{B}\;\Varid{⊗}\;\Conid{C})\;\mathbin{:}\;\Conid{Set}\;\Keyword{where}{}\<[E]%
\\
\>[3]{}\hsindent{2}{}\<[5]%
\>[5]{}\Varid{origin}\;\mathbin{:}\;{}\<[15]%
\>[15]{}(\Varid{h₁}\;{}\<[20]%
\>[20]{}\mathbin{:}\;\mathit{LG}\;\Conid{E}\;\Varid{⊢}\;\Conid{B})\;{}\<[E]%
\\
\>[15]{}(\Varid{h₂}\;{}\<[20]%
\>[20]{}\mathbin{:}\;\mathit{LG}\;\Conid{F}\;\Varid{⊢}\;\Conid{C})\;{}\<[E]%
\\
\>[15]{}(\Varid{f′}\;{}\<[20]%
\>[20]{}\mathbin{:}\;\Varid{∀}\;\{\mskip1.5mu \Conid{G}\mskip1.5mu\}\;\Varid{→}\;\mathit{LG}\;\Conid{E}\;\Varid{⊗}\;\Conid{F}\;\Varid{⊢}\;\Conid{G}\;\Varid{→}\;\mathit{LG}\;\Conid{A}\;\Varid{⊢}\;\Conid{G})\;{}\<[E]%
\\
\>[15]{}(\Varid{pr}\;{}\<[20]%
\>[20]{}\mathbin{:}\;\Varid{f}\;\Varid{≡}\;\Varid{f′}\;(\Varid{m⊗}\;\Varid{h₁}\;\Varid{h₂}))\;{}\<[E]%
\\
\>[20]{}\Varid{→}\;\Conid{Origin}\;\Varid{f}{}\<[E]%
\ColumnHook
\end{hscode}\resethooks
In the above snippet, we have chosen to leave the quantifier \ensuremath{\Varid{∀}\;\{\mskip1.5mu \Conid{G}\mskip1.5mu\}}
explicit to stress that \ensuremath{\Varid{f′}} should work for \textit{any} formula \ensuremath{\Conid{G}},
not only for \ensuremath{\Conid{B}\;\Varid{⊗}\;\Conid{C}}.

All that remains now is to show that for any \ensuremath{\Varid{f}} of type \ensuremath{\mathit{LG}\;\Conid{A}\;\Varid{⊢}\;\Conid{B}\;\Varid{⊗}\;\Conid{C}}, we can construct such a view. We will attempt to do this by
induction on the given derivation.
Note that \ensuremath{\{\ \}0} is the Agda syntax for a proof obligation. For
clarity, I have added the types of the various subproofs \ensuremath{\Varid{f}} in
comments:
\begin{hscode}\SaveRestoreHook
\column{B}{@{}>{\hspre}l<{\hspost}@{}}%
\column{3}{@{}>{\hspre}l<{\hspost}@{}}%
\column{14}{@{}>{\hspre}l<{\hspost}@{}}%
\column{20}{@{}>{\hspre}l<{\hspost}@{}}%
\column{E}{@{}>{\hspre}l<{\hspost}@{}}%
\>[3]{}\Varid{find}\;\mathbin{:}\;(\Varid{f}\;\mathbin{:}\;\mathit{LG}\;\Conid{A}\;\Varid{⊢}\;\Conid{B}\;\Varid{⊗}\;\Conid{C})\;\Varid{→}\;\Conid{Origin}\;\Varid{f}{}\<[E]%
\\
\>[3]{}\Varid{find}\;(\Varid{m⊗}\;{}\<[14]%
\>[14]{}\Varid{f}\;\Varid{g})\;{}\<[20]%
\>[20]{}\mathrel{=}\;\Varid{origin}\;\Varid{f}\;\Varid{g}\;\Varid{id}\;\Varid{refl}{}\<[E]%
\\
\>[3]{}\Varid{find}\;(\Varid{r⇒⊗}\;{}\<[14]%
\>[14]{}\Varid{f})\;{}\<[20]%
\>[20]{}\mathrel{=}\;\{\ \}0\mbox{\onelinecomment  f : LG A₂ ⊢ A₁ ⇒ B ⊗ C}{}\<[E]%
\\
\>[3]{}\Varid{find}\;(\Varid{r⇐⊗}\;{}\<[14]%
\>[14]{}\Varid{f})\;{}\<[20]%
\>[20]{}\mathrel{=}\;\{\ \}1\mbox{\onelinecomment  f : LG A₁ ⊢ B ⊗ C ⇐ A₂}{}\<[E]%
\\
\>[3]{}\Varid{find}\;(\Varid{r⊕⇛}\;{}\<[14]%
\>[14]{}\Varid{f})\;{}\<[20]%
\>[20]{}\mathrel{=}\;\{\ \}2\mbox{\onelinecomment  f : LG A₂ ⊢ A₁ ⊕ B ⊗ C}{}\<[E]%
\\
\>[3]{}\Varid{find}\;(\Varid{r⊕⇚}\;{}\<[14]%
\>[14]{}\Varid{f})\;{}\<[20]%
\>[20]{}\mathrel{=}\;\{\ \}3\mbox{\onelinecomment  f : LG A₁ ⊢ B ⊗ C ⊕ A₂}{}\<[E]%
\ColumnHook
\end{hscode}\resethooks
Alas! While in the first case, where \ensuremath{\Varid{f}} is of the form \ensuremath{\Varid{m⊗}\;\Varid{f}\;\Varid{g}}, we
have found our monotonicity rule, the remaining cases are less kind.
It seems that we have neglected to account for derivations where our
cut formula is temporarily nested within another formula.

We will need some new vocabulary to describe what is going on in the
above example.
We would like to describe contexts which a) can be taken apart using
residuation, and b) when fully taken apart, will leave the nested
formula on the correct side of the turnstile. A natural fit for this
is using polarity:
\begin{hscode}\SaveRestoreHook
\column{B}{@{}>{\hspre}l<{\hspost}@{}}%
\column{3}{@{}>{\hspre}l<{\hspost}@{}}%
\column{E}{@{}>{\hspre}l<{\hspost}@{}}%
\>[3]{}\Keyword{data}\;\Conid{Polarity}\;\mathbin{:}\;\Conid{Set}\;\Keyword{where}\;+\;-\;\mathbin{:}\;\Conid{Polarity}{}\<[E]%
\ColumnHook
\end{hscode}\resethooks
Below we define well-polarised formula and judgement contexts with
exactly one hole. We use a $\triangleleft$ or $\triangleright$ to
denote in which argument the hole is:
\begin{hscode}\SaveRestoreHook
\column{B}{@{}>{\hspre}l<{\hspost}@{}}%
\column{3}{@{}>{\hspre}l<{\hspost}@{}}%
\column{5}{@{}>{\hspre}l<{\hspost}@{}}%
\column{11}{@{}>{\hspre}l<{\hspost}@{}}%
\column{19}{@{}>{\hspre}l<{\hspost}@{}}%
\column{26}{@{}>{\hspre}l<{\hspost}@{}}%
\column{34}{@{}>{\hspre}l<{\hspost}@{}}%
\column{E}{@{}>{\hspre}l<{\hspost}@{}}%
\>[3]{}\Keyword{data}\;\Conid{Context}\;(\Varid{p}\;\mathbin{:}\;\Conid{Polarity})\;\mathbin{:}\;\Conid{Polarity}\;\Varid{→}\;\Conid{Set}\;\Keyword{where}{}\<[E]%
\\[\blanklineskip]%
\>[3]{}\hsindent{2}{}\<[5]%
\>[5]{}\Varid{[]}\;{}\<[11]%
\>[11]{}\mathbin{:}\;\Conid{Context}\;\Varid{p}\;\Varid{p}{}\<[E]%
\\[\blanklineskip]%
\>[3]{}\hsindent{2}{}\<[5]%
\>[5]{}\_\holer{\otimes}\_\;{}\<[11]%
\>[11]{}\mathbin{:}\;\Conid{Type}\;{}\<[19]%
\>[19]{}\Varid{→}\;\Conid{Context}\;\Varid{p}\;+\;{}\<[34]%
\>[34]{}\Varid{→}\;\Conid{Context}\;\Varid{p}\;+{}\<[E]%
\\
\>[3]{}\hsindent{2}{}\<[5]%
\>[5]{}\_\holer{\varbslash}\_\;{}\<[11]%
\>[11]{}\mathbin{:}\;\Conid{Type}\;{}\<[19]%
\>[19]{}\Varid{→}\;\Conid{Context}\;\Varid{p}\;-\;{}\<[34]%
\>[34]{}\Varid{→}\;\Conid{Context}\;\Varid{p}\;-{}\<[E]%
\\
\>[3]{}\hsindent{2}{}\<[5]%
\>[5]{}\_\holer{\varslash}\_\;{}\<[11]%
\>[11]{}\mathbin{:}\;\Conid{Type}\;{}\<[19]%
\>[19]{}\Varid{→}\;\Conid{Context}\;\Varid{p}\;+\;{}\<[34]%
\>[34]{}\Varid{→}\;\Conid{Context}\;\Varid{p}\;-{}\<[E]%
\\[\blanklineskip]%
\>[3]{}\hsindent{2}{}\<[5]%
\>[5]{}\_\holel{\otimes}\_\;{}\<[11]%
\>[11]{}\mathbin{:}\;\Conid{Context}\;\Varid{p}\;+\;{}\<[26]%
\>[26]{}\Varid{→}\;\Conid{Type}\;{}\<[34]%
\>[34]{}\Varid{→}\;\Conid{Context}\;\Varid{p}\;+{}\<[E]%
\\
\>[3]{}\hsindent{2}{}\<[5]%
\>[5]{}\_\holel{\varbslash}\_\;{}\<[11]%
\>[11]{}\mathbin{:}\;\Conid{Context}\;\Varid{p}\;+\;{}\<[26]%
\>[26]{}\Varid{→}\;\Conid{Type}\;{}\<[34]%
\>[34]{}\Varid{→}\;\Conid{Context}\;\Varid{p}\;-{}\<[E]%
\\
\>[3]{}\hsindent{2}{}\<[5]%
\>[5]{}\_\holel{\varslash}\_\;{}\<[11]%
\>[11]{}\mathbin{:}\;\Conid{Context}\;\Varid{p}\;-\;{}\<[26]%
\>[26]{}\Varid{→}\;\Conid{Type}\;{}\<[34]%
\>[34]{}\Varid{→}\;\Conid{Context}\;\Varid{p}\;-{}\<[E]%
\\[\blanklineskip]%
\>[3]{}\hsindent{2}{}\<[5]%
\>[5]{}\_\holer{\oplus}\_\;{}\<[11]%
\>[11]{}\mathbin{:}\;\Conid{Type}\;{}\<[19]%
\>[19]{}\Varid{→}\;\Conid{Context}\;\Varid{p}\;-\;{}\<[34]%
\>[34]{}\Varid{→}\;\Conid{Context}\;\Varid{p}\;-{}\<[E]%
\\
\>[3]{}\hsindent{2}{}\<[5]%
\>[5]{}\_\holer{\varoslash}\_\;{}\<[11]%
\>[11]{}\mathbin{:}\;\Conid{Type}\;{}\<[19]%
\>[19]{}\Varid{→}\;\Conid{Context}\;\Varid{p}\;-\;{}\<[34]%
\>[34]{}\Varid{→}\;\Conid{Context}\;\Varid{p}\;+{}\<[E]%
\\
\>[3]{}\hsindent{2}{}\<[5]%
\>[5]{}\_\holer{\varobslash}\_\;{}\<[11]%
\>[11]{}\mathbin{:}\;\Conid{Type}\;{}\<[19]%
\>[19]{}\Varid{→}\;\Conid{Context}\;\Varid{p}\;+\;{}\<[34]%
\>[34]{}\Varid{→}\;\Conid{Context}\;\Varid{p}\;+{}\<[E]%
\\[\blanklineskip]%
\>[3]{}\hsindent{2}{}\<[5]%
\>[5]{}\_\holel{\oplus}\_\;{}\<[11]%
\>[11]{}\mathbin{:}\;\Conid{Context}\;\Varid{p}\;-\;{}\<[26]%
\>[26]{}\Varid{→}\;\Conid{Type}\;{}\<[34]%
\>[34]{}\Varid{→}\;\Conid{Context}\;\Varid{p}\;-{}\<[E]%
\\
\>[3]{}\hsindent{2}{}\<[5]%
\>[5]{}\_\holel{\varoslash}\_\;{}\<[11]%
\>[11]{}\mathbin{:}\;\Conid{Context}\;\Varid{p}\;+\;{}\<[26]%
\>[26]{}\Varid{→}\;\Conid{Type}\;{}\<[34]%
\>[34]{}\Varid{→}\;\Conid{Context}\;\Varid{p}\;+{}\<[E]%
\\
\>[3]{}\hsindent{2}{}\<[5]%
\>[5]{}\_\holel{\varobslash}\_\;{}\<[11]%
\>[11]{}\mathbin{:}\;\Conid{Context}\;\Varid{p}\;-\;{}\<[26]%
\>[26]{}\Varid{→}\;\Conid{Type}\;{}\<[34]%
\>[34]{}\Varid{→}\;\Conid{Context}\;\Varid{p}\;+{}\<[E]%
\\[\blanklineskip]%
\>[3]{}\Keyword{data}\;\Conid{Contextᴶ}\;(\Varid{p}\;\mathbin{:}\;\Conid{Polarity})\;\mathbin{:}\;\Conid{Set}\;\Keyword{where}{}\<[E]%
\\[\blanklineskip]%
\>[3]{}\hsindent{2}{}\<[5]%
\>[5]{}\_\holel{\vdash}\_\;{}\<[11]%
\>[11]{}\mathbin{:}\;\Conid{Context}\;\Varid{p}\;+\;{}\<[26]%
\>[26]{}\Varid{→}\;\Conid{Type}\;{}\<[34]%
\>[34]{}\Varid{→}\;\Conid{Contextᴶ}\;\Varid{p}{}\<[E]%
\\
\>[3]{}\hsindent{2}{}\<[5]%
\>[5]{}\_\holer{\vdash}\_\;{}\<[11]%
\>[11]{}\mathbin{:}\;\Conid{Type}\;{}\<[19]%
\>[19]{}\Varid{→}\;\Conid{Context}\;\Varid{p}\;-\;{}\<[34]%
\>[34]{}\Varid{→}\;\Conid{Contextᴶ}\;\Varid{p}{}\<[E]%
\ColumnHook
\end{hscode}\resethooks
We also define two operators which, given a context and a formula,
will fill the hole in the given context with the given formula. The
definition for \ensuremath{\Varid{\char95 [\char95 ]}} is entirely predictable and repetitive, and has
been mostly omitted\footnote{%
  For the remainder of this paper, any partial omission of a function
  will be denoted with an ellipsis at the end of the code block.
}:
\setboolean{partial}{true}%
\begin{hscode}\SaveRestoreHook
\column{B}{@{}>{\hspre}l<{\hspost}@{}}%
\column{3}{@{}>{\hspre}l<{\hspost}@{}}%
\column{13}{@{}>{\hspre}l<{\hspost}@{}}%
\column{E}{@{}>{\hspre}l<{\hspost}@{}}%
\>[3]{}\Varid{\char95 [\char95 ]}\;\mathbin{:}\;\Conid{Context}\;\Varid{p₁}\;\Varid{p₂}\;\Varid{→}\;\Conid{Type}\;\Varid{→}\;\Conid{Type}{}\<[E]%
\\
\>[3]{}\Varid{[]}\;{}\<[13]%
\>[13]{}[\mskip1.5mu \;\Conid{A}\;\mskip1.5mu]\;\mathrel{=}\;\Conid{A}{}\<[E]%
\\
\>[3]{}(\Conid{B}\;\holer{\otimes}\;\Conid{C})\;{}\<[13]%
\>[13]{}[\mskip1.5mu \;\Conid{A}\;\mskip1.5mu]\;\mathrel{=}\;\Conid{B}\;\Varid{⊗}\;(\Conid{C}\;[\mskip1.5mu \;\Conid{A}\;\mskip1.5mu]){}\<[E]%
\ColumnHook
\end{hscode}\resethooks
\setboolean{partial}{false}%
\begin{hscode}\SaveRestoreHook
\column{B}{@{}>{\hspre}l<{\hspost}@{}}%
\column{3}{@{}>{\hspre}l<{\hspost}@{}}%
\column{E}{@{}>{\hspre}l<{\hspost}@{}}%
\>[3]{}\Varid{\char95 [\char95 ]ᴶ}\;\mathbin{:}\;\Conid{Contextᴶ}\;\Varid{p}\;\Varid{→}\;\Conid{Type}\;\Varid{→}\;\Conid{Judgement}{}\<[E]%
\\
\>[3]{}(\Conid{A}\;\holel{\vdash}\;\Conid{B})\;[\mskip1.5mu \;\Conid{C}\;\Varid{]ᴶ}\;\mathrel{=}\;\Conid{A}\;[\mskip1.5mu \;\Conid{C}\;\mskip1.5mu]\;\Varid{⊢}\;\Conid{B}{}\<[E]%
\\
\>[3]{}(\Conid{A}\;\holer{\vdash}\;\Conid{B})\;[\mskip1.5mu \;\Conid{C}\;\Varid{]ᴶ}\;\mathrel{=}\;\Conid{A}\;\Varid{⊢}\;\Conid{B}\;[\mskip1.5mu \;\Conid{C}\;\mskip1.5mu]{}\<[E]%
\ColumnHook
\end{hscode}\resethooks
The crucial point about these well-polarised judgement contexts is
that, once the entire context is peeled away, the formula will be
at the top level on the side corresponding to the polarity
argument---with $+$ and $-$ corresponding to the antecedent and the
succedent, respectively. Therefore, in order to generalise our
previous definition of \ensuremath{\Conid{Origin}}, we want the occurrence of \ensuremath{\Conid{B}\;\Varid{⊗}\;\Conid{C}} to
be nested in a \textit{negative} context:
\begin{hscode}\SaveRestoreHook
\column{B}{@{}>{\hspre}l<{\hspost}@{}}%
\column{5}{@{}>{\hspre}l<{\hspost}@{}}%
\column{7}{@{}>{\hspre}l<{\hspost}@{}}%
\column{18}{@{}>{\hspre}l<{\hspost}@{}}%
\column{21}{@{}>{\hspre}l<{\hspost}@{}}%
\column{23}{@{}>{\hspre}l<{\hspost}@{}}%
\column{24}{@{}>{\hspre}l<{\hspost}@{}}%
\column{E}{@{}>{\hspre}l<{\hspost}@{}}%
\>[5]{}\Keyword{data}\;\Conid{Origin′}\;(\Conid{J}\;{}\<[24]%
\>[24]{}\mathbin{:}\;\Conid{Contextᴶ}\;-)\;{}\<[E]%
\\
\>[5]{}\hsindent{13}{}\<[18]%
\>[18]{}(\Varid{f}\;{}\<[23]%
\>[23]{}\mathbin{:}\;\mathit{LG}\;\Conid{J}\;[\mskip1.5mu \;\Conid{B}\;\Varid{⊗}\;\Conid{C}\;\Varid{]ᴶ})\;{}\<[E]%
\\
\>[23]{}\mathbin{:}\;\Conid{Set}\;\Keyword{where}{}\<[E]%
\\[\blanklineskip]%
\>[5]{}\hsindent{2}{}\<[7]%
\>[7]{}\Varid{origin}\;\mathbin{:}\;(\Varid{h₁}\;{}\<[21]%
\>[21]{}\mathbin{:}\;\mathit{LG}\;\Conid{E}\;\Varid{⊢}\;\Conid{B})\;{}\<[E]%
\\
\>[7]{}\hsindent{11}{}\<[18]%
\>[18]{}(\Varid{h₂}\;{}\<[23]%
\>[23]{}\mathbin{:}\;\mathit{LG}\;\Conid{F}\;\Varid{⊢}\;\Conid{C})\;{}\<[E]%
\\
\>[7]{}\hsindent{11}{}\<[18]%
\>[18]{}(\Varid{f′}\;{}\<[23]%
\>[23]{}\mathbin{:}\;\mathit{LG}\;\Conid{E}\;\Varid{⊗}\;\Conid{F}\;\Varid{⊢}\;\Conid{G}\;\Varid{→}\;\mathit{LG}\;\Conid{J}\;[\mskip1.5mu \;\Conid{G}\;\Varid{]ᴶ})\;{}\<[E]%
\\
\>[7]{}\hsindent{11}{}\<[18]%
\>[18]{}(\Varid{pr}\;{}\<[23]%
\>[23]{}\mathbin{:}\;\Varid{f}\;\Varid{≡}\;\Varid{f′}\;(\Varid{m⊗}\;\Varid{h₁}\;\Varid{h₂}))\;{}\<[E]%
\\
\>[23]{}\Varid{→}\;\Conid{Origin′}\;\Conid{J}\;\Varid{f}{}\<[E]%
\ColumnHook
\end{hscode}\resethooks
Using this more general definition \ensuremath{\Conid{Origin′}}, we can define a more general
function \ensuremath{\Varid{find′}}---and this time, our proof by induction works!

Note that in Agda, the \ensuremath{\Keyword{with}} construct is used to pattern match on
the result of an expression:
\setboolean{partial}{true}%
\begin{hscode}\SaveRestoreHook
\column{B}{@{}>{\hspre}l<{\hspost}@{}}%
\column{7}{@{}>{\hspre}l<{\hspost}@{}}%
\column{30}{@{}>{\hspre}l<{\hspost}@{}}%
\column{35}{@{}>{\hspre}l<{\hspost}@{}}%
\column{40}{@{}>{\hspre}l<{\hspost}@{}}%
\column{E}{@{}>{\hspre}l<{\hspost}@{}}%
\>[7]{}\Varid{find′}\;\mathbin{:}\;(\Conid{J}\;\mathbin{:}\;\Conid{Contextᴶ}\;-)\;(\Varid{f}\;\mathbin{:}\;\mathit{LG}\;\Conid{J}\;[\mskip1.5mu \;\Conid{B}\;\Varid{⊗}\;\Conid{C}\;\Varid{]ᴶ})\;\Varid{→}\;\Conid{Origin′}\;\Conid{J}\;\Varid{f}{}\<[E]%
\\
\>[7]{}\Varid{find′}\;(\Varid{.\char95 }\;\holer{\vdash}\;\Varid{[]})\;{}\<[30]%
\>[30]{}(\Varid{m⊗}\;{}\<[35]%
\>[35]{}\Varid{f}\;\Varid{g})\;\mathrel{=}\;\Varid{origin}\;\Varid{f}\;\Varid{g}\;\Varid{id}\;\Varid{refl}{}\<[E]%
\\
\>[7]{}\Varid{find′}\;(\Varid{.\char95 }\;\holer{\vdash}\;\Varid{[]})\;{}\<[30]%
\>[30]{}(\Varid{r⇒⊗}\;\Varid{f})\;{}\<[40]%
\>[40]{}\Keyword{with}\;\Varid{find′}\;(\anonymous \;\holer{\vdash}\;(\anonymous \;\holer{\varbslash}\;\Varid{[]}))\;\Varid{f}{}\<[E]%
\\
\>[7]{}\Varid{...}\;\mid \;\Varid{origin}\;\Varid{h₁}\;\Varid{h₂}\;\Varid{f′}\;\Varid{pr}\;\Varid{rewrite}\;\Varid{pr}\;\mathrel{=}\;\Varid{origin}\;\Varid{h₁}\;\Varid{h₂}\;(\Varid{r⇒⊗}\;\Varid{∘}\;\Varid{f′})\;\Varid{refl}{}\<[E]%
\ColumnHook
\end{hscode}\resethooks
\setboolean{partial}{false}%
However, there are many cases---97 in total. The reason for this is
that the possible derivation steps depend on the main connective;
therefore we first have to explore every possible main connective, and
then every possible rule which would produce that main connective.
Because of this, the definitions of the various \ensuremath{\Varid{find′}} functions are
very long and tedious, and have mostly been omitted.\footnote{%
  The burden on the programmer or logician can be reduced by clever
  use of the symmetries $\cdot^{\bowtie}$ and $\cdot^{\infty}$ as done
  in \citet{moortgat2009}. One would have to implement only
  \textit{three} of the \ensuremath{\Varid{find′}} functions (e.g. for \ensuremath{\Varid{el}}, \ensuremath{\Varid{⊗}} and
  \ensuremath{\Varid{⇒}}); the remaining four can then be derived using the symmetries.
}

From the more general \ensuremath{\Conid{Origin′}} and \ensuremath{\Varid{find′}} we can very easily recover
our original definitions \ensuremath{\Conid{Origin}} and \ensuremath{\Varid{find}} by setting the context
to be empty. In the case of the cut formula \ensuremath{\Conid{B}\;\Varid{⊗}\;\Conid{C}}, we set the
context to \ensuremath{(\anonymous \;\holer{\vdash}\;\Varid{[]})} to ensure that the formula ends up at the top
level in the succedent:
\begin{hscode}\SaveRestoreHook
\column{B}{@{}>{\hspre}l<{\hspost}@{}}%
\column{7}{@{}>{\hspre}l<{\hspost}@{}}%
\column{15}{@{}>{\hspre}l<{\hspost}@{}}%
\column{28}{@{}>{\hspre}l<{\hspost}@{}}%
\column{E}{@{}>{\hspre}l<{\hspost}@{}}%
\>[7]{}\Conid{Origin}\;{}\<[15]%
\>[15]{}\mathbin{:}\;(\Varid{f}\;\mathbin{:}\;\mathit{LG}\;\Conid{A}\;\Varid{⊢}\;\Conid{B}\;\Varid{⊗}\;\Conid{C})\;\Varid{→}\;\Conid{Set}{}\<[E]%
\\
\>[7]{}\Conid{Origin}\;{}\<[15]%
\>[15]{}\Varid{f}\;\mathrel{=}\;\Conid{Origin′}\;{}\<[28]%
\>[28]{}(\anonymous \;\holer{\vdash}\;\Varid{[]})\;\Varid{f}{}\<[E]%
\\[\blanklineskip]%
\>[7]{}\Varid{find}\;{}\<[15]%
\>[15]{}\mathbin{:}\;(\Varid{f}\;\mathbin{:}\;\mathit{LG}\;\Conid{A}\;\Varid{⊢}\;\Conid{B}\;\Varid{⊗}\;\Conid{C})\;\Varid{→}\;\Conid{Origin}\;\Varid{f}{}\<[E]%
\\
\>[7]{}\Varid{find}\;{}\<[15]%
\>[15]{}\Varid{f}\;\mathrel{=}\;\Varid{find′}\;{}\<[28]%
\>[28]{}(\anonymous \;\holer{\vdash}\;\Varid{[]})\;\Varid{f}{}\<[E]%
\ColumnHook
\end{hscode}\resethooks
And with that, we can finally put the rewrite rules from
\citet{moortgat1999} to use. We can define \ensuremath{\Varid{cut′}} by pattern
matching on the cut formula \ensuremath{\Conid{B}}; applying the appropriate \ensuremath{\Varid{find′}}
function to find′ the monotonicity rule introducing the formula; and
apply the appropriate rewrite rule to create a derivation containing
two cuts on structurally smaller formulas:
\begin{hscode}\SaveRestoreHook
\column{B}{@{}>{\hspre}l<{\hspost}@{}}%
\column{3}{@{}>{\hspre}l<{\hspost}@{}}%
\column{8}{@{}>{\hspre}l<{\hspost}@{}}%
\column{21}{@{}>{\hspre}l<{\hspost}@{}}%
\column{24}{@{}>{\hspre}l<{\hspost}@{}}%
\column{28}{@{}>{\hspre}l<{\hspost}@{}}%
\column{35}{@{}>{\hspre}l<{\hspost}@{}}%
\column{E}{@{}>{\hspre}l<{\hspost}@{}}%
\>[3]{}\Varid{cut′}\;\mathbin{:}\;(\Varid{f}\;\mathbin{:}\;\mathit{LG}\;\Conid{A}\;\Varid{⊢}\;\Conid{B})\;(\Varid{g}\;\mathbin{:}\;\mathit{LG}\;\Conid{B}\;\Varid{⊢}\;\Conid{C})\;\Varid{→}\;\mathit{LG}\;\Conid{A}\;\Varid{⊢}\;\Conid{C}{}\<[E]%
\\
\>[3]{}\Varid{cut′}\;\{\mskip1.5mu \Conid{B}\;\mathrel{=}\;\Varid{el}\;\anonymous \mskip1.5mu\}\;{}\<[21]%
\>[21]{}\Varid{f}\;{}\<[24]%
\>[24]{}\Varid{g}\;\Keyword{with}\;\Varid{el.find}\;\Varid{g}{}\<[E]%
\\
\>[3]{}\Varid{...}\;{}\<[8]%
\>[8]{}\mid \;(\Varid{el.origin}\;{}\<[28]%
\>[28]{}\Varid{g′}\;\anonymous )\;{}\<[35]%
\>[35]{}\mathrel{=}\;\Varid{g′}\;\Varid{f}{}\<[E]%
\\
\>[3]{}\Varid{cut′}\;\{\mskip1.5mu \Conid{B}\;\mathrel{=}\;\anonymous \;\Varid{⊗}\;\anonymous \mskip1.5mu\}\;{}\<[21]%
\>[21]{}\Varid{f}\;{}\<[24]%
\>[24]{}\Varid{g}\;\Keyword{with}\;\Varid{⊗.find}\;\Varid{f}{}\<[E]%
\\
\>[3]{}\Varid{...}\;{}\<[8]%
\>[8]{}\mid \;(\Varid{⊗.origin}\;{}\<[21]%
\>[21]{}\Varid{h₁}\;\Varid{h₂}\;{}\<[28]%
\>[28]{}\Varid{f′}\;\anonymous )\;{}\<[E]%
\\
\>[8]{}\mathrel{=}\;\Varid{f′}\;(\Varid{r⇐⊗}\;(\Varid{cut′}\;\Varid{h₁}\;(\Varid{r⊗⇐}\;(\Varid{r⇒⊗}\;(\Varid{cut′}\;\Varid{h₂}\;(\Varid{r⊗⇒}\;\Varid{g})))))){}\<[E]%
\\
\>[3]{}\Varid{cut′}\;\{\mskip1.5mu \Conid{B}\;\mathrel{=}\;\anonymous \;\Varid{⇐}\;\anonymous \mskip1.5mu\}\;{}\<[21]%
\>[21]{}\Varid{f}\;{}\<[24]%
\>[24]{}\Varid{g}\;\Keyword{with}\;\Varid{⇐.find}\;\Varid{g}{}\<[E]%
\\
\>[3]{}\Varid{...}\;{}\<[8]%
\>[8]{}\mid \;(\Varid{⇐.origin}\;{}\<[21]%
\>[21]{}\Varid{h₁}\;\Varid{h₂}\;{}\<[28]%
\>[28]{}\Varid{g′}\;\anonymous )\;{}\<[E]%
\\
\>[8]{}\mathrel{=}\;\Varid{g′}\;(\Varid{r⊗⇐}\;(\Varid{r⇒⊗}\;(\Varid{cut′}\;\Varid{h₂}\;(\Varid{r⊗⇒}\;(\Varid{cut′}\;(\Varid{r⇐⊗}\;\Varid{f})\;\Varid{h₁}))))){}\<[E]%
\\
\>[3]{}\Varid{cut′}\;\{\mskip1.5mu \Conid{B}\;\mathrel{=}\;\anonymous \;\Varid{⇒}\;\anonymous \mskip1.5mu\}\;{}\<[21]%
\>[21]{}\Varid{f}\;{}\<[24]%
\>[24]{}\Varid{g}\;\Keyword{with}\;\Varid{⇒.find}\;\Varid{g}{}\<[E]%
\\
\>[3]{}\Varid{...}\;{}\<[8]%
\>[8]{}\mid \;(\Varid{⇒.origin}\;{}\<[21]%
\>[21]{}\Varid{h₁}\;\Varid{h₂}\;{}\<[28]%
\>[28]{}\Varid{g′}\;\anonymous )\;{}\<[E]%
\\
\>[8]{}\mathrel{=}\;\Varid{g′}\;(\Varid{r⊗⇒}\;(\Varid{r⇐⊗}\;(\Varid{cut′}\;\Varid{h₁}\;(\Varid{r⊗⇐}\;(\Varid{cut′}\;(\Varid{r⇒⊗}\;\Varid{f})\;\Varid{h₂}))))){}\<[E]%
\\
\>[3]{}\Varid{cut′}\;\{\mskip1.5mu \Conid{B}\;\mathrel{=}\;\anonymous \;\Varid{⊕}\;\anonymous \mskip1.5mu\}\;{}\<[21]%
\>[21]{}\Varid{f}\;{}\<[24]%
\>[24]{}\Varid{g}\;\Keyword{with}\;\Varid{⊕.find}\;\Varid{g}{}\<[E]%
\\
\>[3]{}\Varid{...}\;{}\<[8]%
\>[8]{}\mid \;(\Varid{⊕.origin}\;{}\<[21]%
\>[21]{}\Varid{h₁}\;\Varid{h₂}\;{}\<[28]%
\>[28]{}\Varid{g′}\;\anonymous )\;{}\<[E]%
\\
\>[8]{}\mathrel{=}\;\Varid{g′}\;(\Varid{r⇚⊕}\;(\Varid{cut′}\;(\Varid{r⊕⇚}\;(\Varid{r⇛⊕}\;(\Varid{cut′}\;(\Varid{r⊕⇛}\;\Varid{f})\;\Varid{h₂})))\;\Varid{h₁})){}\<[E]%
\\
\>[3]{}\Varid{cut′}\;\{\mskip1.5mu \Conid{B}\;\mathrel{=}\;\anonymous \;\Varid{⇚}\;\anonymous \mskip1.5mu\}\;{}\<[21]%
\>[21]{}\Varid{f}\;{}\<[24]%
\>[24]{}\Varid{g}\;\Keyword{with}\;\Varid{⇚.find}\;\Varid{f}{}\<[E]%
\\
\>[3]{}\Varid{...}\;{}\<[8]%
\>[8]{}\mid \;(\Varid{⇚.origin}\;{}\<[21]%
\>[21]{}\Varid{h₁}\;\Varid{h₂}\;{}\<[28]%
\>[28]{}\Varid{f′}\;\anonymous )\;{}\<[E]%
\\
\>[8]{}\mathrel{=}\;\Varid{f′}\;(\Varid{r⊕⇚}\;(\Varid{r⇛⊕}\;(\Varid{cut′}\;(\Varid{r⊕⇛}\;(\Varid{cut′}\;\Varid{h₁}\;(\Varid{r⇚⊕}\;\Varid{g})))\;\Varid{h₂}))){}\<[E]%
\\
\>[3]{}\Varid{cut′}\;\{\mskip1.5mu \Conid{B}\;\mathrel{=}\;\anonymous \;\Varid{⇛}\;\anonymous \mskip1.5mu\}\;{}\<[21]%
\>[21]{}\Varid{f}\;{}\<[24]%
\>[24]{}\Varid{g}\;\Keyword{with}\;\Varid{⇛.find}\;\Varid{f}{}\<[E]%
\\
\>[3]{}\Varid{...}\;{}\<[8]%
\>[8]{}\mid \;(\Varid{⇛.origin}\;{}\<[21]%
\>[21]{}\Varid{h₁}\;\Varid{h₂}\;{}\<[28]%
\>[28]{}\Varid{f′}\;\anonymous )\;{}\<[E]%
\\
\>[8]{}\mathrel{=}\;\Varid{f′}\;(\Varid{r⊕⇛}\;(\Varid{r⇚⊕}\;(\Varid{cut′}\;(\Varid{r⊕⇚}\;(\Varid{cut′}\;\Varid{h₂}\;(\Varid{r⇛⊕}\;\Varid{g})))\;\Varid{h₁}))){}\<[E]%
\ColumnHook
\end{hscode}\resethooks

\section{CPS Translation}

For this paper, we have opted to implement the call-by-value CPS
translation as described in \citet{moortgat2007}. This translation
consists of three elements:
\begin{itemize}
\item%
  a function \ensuremath{\Varid{⌈\char95 ⌉}}, which translates formulas in LG to formulas in the
  target system---while we have chosen to translate to Agda, the
  original translation targeted multiplicative intuitionistic linear
  logic;
\item%
  a pair of mutually recursive functions \ensuremath{\Varid{⌈\char95 ⌉ᴸ}} and \ensuremath{\Varid{⌈\char95 ⌉ᴿ}}, which
  translate terms in LG to terms in the target system.
\end{itemize}
In order to write these functions, we will need two additional pieces
of information: a function \ensuremath{\Varid{⌈\char95 ⌉ᵁ}}, which translates the atomic
formulas to Agda types; and a return type \ensuremath{\Conid{R}}, which we will use to
define a ``negation'' as \ensuremath{\Varid{¬}\;\Conid{A}\;\mathrel{=}\;\Conid{A}\;\Varid{→}\;\Conid{R}}. We will therefore implement
the CPS translation in a sub-module, which abstracts over these terms:
\begin{hscode}\SaveRestoreHook
\column{B}{@{}>{\hspre}l<{\hspost}@{}}%
\column{3}{@{}>{\hspre}l<{\hspost}@{}}%
\column{E}{@{}>{\hspre}l<{\hspost}@{}}%
\>[3]{}\Keyword{module}\;\Varid{translation}\;(\Varid{⌈\char95 ⌉ᵁ}\;\mathbin{:}\;\Conid{Univ}\;\Varid{→}\;\Conid{Set})\;(\Conid{R}\;\mathbin{:}\;\Conid{Set})\;\Keyword{where}{}\<[E]%
\ColumnHook
\end{hscode}\resethooks
When using this module, we will generally identify the return type
\ensuremath{\Conid{R}} with the type \ensuremath{\Conid{Bool}} for booleans. However, abstracting over it
will ensure that we do not accidentally use this knowledge during the
translation.

The type-level translation itself maps formulas in LG to types in
Agda, as follows:
\begin{hscode}\SaveRestoreHook
\column{B}{@{}>{\hspre}l<{\hspost}@{}}%
\column{5}{@{}>{\hspre}l<{\hspost}@{}}%
\column{12}{@{}>{\hspre}l<{\hspost}@{}}%
\column{15}{@{}>{\hspre}l<{\hspost}@{}}%
\column{22}{@{}>{\hspre}l<{\hspost}@{}}%
\column{26}{@{}>{\hspre}l<{\hspost}@{}}%
\column{37}{@{}>{\hspre}l<{\hspost}@{}}%
\column{E}{@{}>{\hspre}l<{\hspost}@{}}%
\>[5]{}\Varid{⌈\char95 ⌉}\;\mathbin{:}\;\Conid{Type}\;\Varid{→}\;\Conid{Set}{}\<[E]%
\\
\>[5]{}\Varid{⌈}\;\Varid{el}\;{}\<[12]%
\>[12]{}\Conid{A}\;{}\<[15]%
\>[15]{}\Varid{⌉}\;\mathrel{=}\;{}\<[26]%
\>[26]{}\Varid{⌈}\;\Conid{A}\;\Varid{⌉ᵁ}{}\<[E]%
\\
\>[5]{}\Varid{⌈}\;\Conid{A}\;\Varid{⊗}\;{}\<[12]%
\>[12]{}\Conid{B}\;{}\<[15]%
\>[15]{}\Varid{⌉}\;\mathrel{=}\;{}\<[22]%
\>[22]{}({}\<[26]%
\>[26]{}\Varid{⌈}\;\Conid{A}\;\Varid{⌉}\;\Varid{×}\;{}\<[37]%
\>[37]{}\Varid{⌈}\;\Conid{B}\;\Varid{⌉}){}\<[E]%
\\
\>[5]{}\Varid{⌈}\;\Conid{A}\;\Varid{⇒}\;{}\<[12]%
\>[12]{}\Conid{B}\;{}\<[15]%
\>[15]{}\Varid{⌉}\;\mathrel{=}\;\Varid{¬}\;{}\<[22]%
\>[22]{}({}\<[26]%
\>[26]{}\Varid{⌈}\;\Conid{A}\;\Varid{⌉}\;\Varid{×}\;\Varid{¬}\;{}\<[37]%
\>[37]{}\Varid{⌈}\;\Conid{B}\;\Varid{⌉}){}\<[E]%
\\
\>[5]{}\Varid{⌈}\;\Conid{B}\;\Varid{⇐}\;{}\<[12]%
\>[12]{}\Conid{A}\;{}\<[15]%
\>[15]{}\Varid{⌉}\;\mathrel{=}\;\Varid{¬}\;{}\<[22]%
\>[22]{}(\Varid{¬}\;{}\<[26]%
\>[26]{}\Varid{⌈}\;\Conid{B}\;\Varid{⌉}\;\Varid{×}\;{}\<[37]%
\>[37]{}\Varid{⌈}\;\Conid{A}\;\Varid{⌉}){}\<[E]%
\\
\>[5]{}\Varid{⌈}\;\Conid{B}\;\Varid{⊕}\;{}\<[12]%
\>[12]{}\Conid{A}\;{}\<[15]%
\>[15]{}\Varid{⌉}\;\mathrel{=}\;\Varid{¬}\;{}\<[22]%
\>[22]{}(\Varid{¬}\;{}\<[26]%
\>[26]{}\Varid{⌈}\;\Conid{B}\;\Varid{⌉}\;\Varid{×}\;\Varid{¬}\;{}\<[37]%
\>[37]{}\Varid{⌈}\;\Conid{A}\;\Varid{⌉}){}\<[E]%
\\
\>[5]{}\Varid{⌈}\;\Conid{B}\;\Varid{⇚}\;{}\<[12]%
\>[12]{}\Conid{A}\;{}\<[15]%
\>[15]{}\Varid{⌉}\;\mathrel{=}\;{}\<[22]%
\>[22]{}({}\<[26]%
\>[26]{}\Varid{⌈}\;\Conid{B}\;\Varid{⌉}\;\Varid{×}\;\Varid{¬}\;{}\<[37]%
\>[37]{}\Varid{⌈}\;\Conid{A}\;\Varid{⌉}){}\<[E]%
\\
\>[5]{}\Varid{⌈}\;\Conid{A}\;\Varid{⇛}\;{}\<[12]%
\>[12]{}\Conid{B}\;{}\<[15]%
\>[15]{}\Varid{⌉}\;\mathrel{=}\;{}\<[22]%
\>[22]{}(\Varid{¬}\;{}\<[26]%
\>[26]{}\Varid{⌈}\;\Conid{A}\;\Varid{⌉}\;\Varid{×}\;{}\<[37]%
\>[37]{}\Varid{⌈}\;\Conid{B}\;\Varid{⌉}){}\<[E]%
\ColumnHook
\end{hscode}\resethooks
The translations on terms map terms in LG to the Agda function
space. Each LG term is associated with \textit{two} functions,
depending on whether the focus is on $A$ or $B$ as the active formula:
\setboolean{partial}{true}%
\begin{hscode}\SaveRestoreHook
\column{B}{@{}>{\hspre}l<{\hspost}@{}}%
\column{5}{@{}>{\hspre}l<{\hspost}@{}}%
\column{7}{@{}>{\hspre}l<{\hspost}@{}}%
\column{E}{@{}>{\hspre}l<{\hspost}@{}}%
\>[5]{}\Keyword{mutual}{}\<[E]%
\\
\>[5]{}\hsindent{2}{}\<[7]%
\>[7]{}\Varid{⌈\char95 ⌉ᴸ}\;\mathbin{:}\;\mathit{LG}\;\Conid{A}\;\Varid{⊢}\;\Conid{B}\;\Varid{→}\;\Varid{¬}\;\Varid{⌈}\;\Conid{B}\;\Varid{⌉}\;\Varid{→}\;\Varid{¬}\;\Varid{⌈}\;\Conid{A}\;\Varid{⌉}{}\<[E]%
\\
\>[5]{}\hsindent{2}{}\<[7]%
\>[7]{}\Varid{⌈\char95 ⌉ᴿ}\;\mathbin{:}\;\mathit{LG}\;\Conid{A}\;\Varid{⊢}\;\Conid{B}\;\Varid{→}\;\Varid{⌈}\;\Conid{A}\;\Varid{⌉}\;\Varid{→}\;\Varid{¬}\;\Varid{¬}\;\Varid{⌈}\;\Conid{B}\;\Varid{⌉}{}\<[E]%
\ColumnHook
\end{hscode}\resethooks
\setboolean{partial}{false}%
The CPS translations of the terms are rather verbose, and trivial to
deduce, when guided by the translation on types. Therefore, in the
interest of space they have been omitted from the paper.\footnote{%
  They are, however, present in the source and therefore available on
  GitHub.
}

\section{Example}

In this final section, we will present the analysis of an example
sentence, using the type-logical grammar implemented above. The
example we will analyse is:
\begin{quote}
``Someone loves everyone.''
\end{quote}
This sentence is well known to be ambiguous, owing to the presence of
the two quantifiers. There are two readings:
\begin{enumerate}
\item[a.] There is some person who loves every person.
\item[b.] For each person, there is some person who loves them.
\end{enumerate}
We will demonstrate that the system, as implemented in this paper,
accurately captures these readings.

Before we can do that, however, there is a small amount of boiler
plate that we have to deal with: we still need to choose a
representation for our atomic types, and show how these translate into
Agda.
In what follows, we will assume we have access to a type for entities,
suitable definitions for the universal and existential quantifiers,
and meanings for `loves' and `person':
\begin{hscode}\SaveRestoreHook
\column{B}{@{}>{\hspre}l<{\hspost}@{}}%
\column{3}{@{}>{\hspre}l<{\hspost}@{}}%
\column{5}{@{}>{\hspre}l<{\hspost}@{}}%
\column{13}{@{}>{\hspre}l<{\hspost}@{}}%
\column{E}{@{}>{\hspre}l<{\hspost}@{}}%
\>[3]{}\Keyword{postulate}{}\<[E]%
\\
\>[3]{}\hsindent{2}{}\<[5]%
\>[5]{}\Conid{Entity}\;{}\<[13]%
\>[13]{}\mathbin{:}\;\Conid{Set}{}\<[E]%
\\
\>[3]{}\hsindent{2}{}\<[5]%
\>[5]{}\forall\;{}\<[13]%
\>[13]{}\mathbin{:}\;(\Conid{Entity}\;\Varid{→}\;\Conid{Bool})\;\Varid{→}\;\Conid{Bool}{}\<[E]%
\\
\>[3]{}\hsindent{2}{}\<[5]%
\>[5]{}\exists\;{}\<[13]%
\>[13]{}\mathbin{:}\;(\Conid{Entity}\;\Varid{→}\;\Conid{Bool})\;\Varid{→}\;\Conid{Bool}{}\<[E]%
\\
\>[3]{}\hsindent{2}{}\<[5]%
\>[5]{}\textsc{loves}\;{}\<[13]%
\>[13]{}\mathbin{:}\;\Conid{Entity}\;\Varid{→}\;\Conid{Entity}\;\Varid{→}\;\Conid{Bool}{}\<[E]%
\\
\>[3]{}\hsindent{2}{}\<[5]%
\>[5]{}\textsc{person}\;{}\<[13]%
\>[13]{}\mathbin{:}\;\Conid{Entity}\;\Varid{→}\;\Conid{Bool}{}\<[E]%
\ColumnHook
\end{hscode}\resethooks
We will instantiate the type for atomic formulas to \ensuremath{\Conid{Univ}}, as defined
below:
\begin{hscode}\SaveRestoreHook
\column{B}{@{}>{\hspre}l<{\hspost}@{}}%
\column{3}{@{}>{\hspre}l<{\hspost}@{}}%
\column{E}{@{}>{\hspre}l<{\hspost}@{}}%
\>[3]{}\Keyword{data}\;\Conid{Univ}\;\mathbin{:}\;\Conid{Set}\;\Keyword{where}\;\Conid{N}\;\Conid{NP}\;\Conid{S}\;\mathbin{:}\;\Conid{Univ}{}\<[E]%
\ColumnHook
\end{hscode}\resethooks
Last, we need to define a function which maps the values of \ensuremath{\Conid{Univ}} to
Agda types. We would like to map the atomic formulas as follows:
\begin{hscode}\SaveRestoreHook
\column{B}{@{}>{\hspre}l<{\hspost}@{}}%
\column{3}{@{}>{\hspre}l<{\hspost}@{}}%
\column{9}{@{}>{\hspre}l<{\hspost}@{}}%
\column{E}{@{}>{\hspre}l<{\hspost}@{}}%
\>[3]{}\Varid{⌈\char95 ⌉ᵁ}\;\mathbin{:}\;\Conid{Univ}\;\Varid{→}\;\Conid{Set}{}\<[E]%
\\
\>[3]{}\Varid{⌈}\;\Conid{N}\;{}\<[9]%
\>[9]{}\Varid{⌉ᵁ}\;\mathrel{=}\;\Conid{Entity}\;\Varid{→}\;\Conid{Bool}{}\<[E]%
\\
\>[3]{}\Varid{⌈}\;\Conid{NP}\;{}\<[9]%
\>[9]{}\Varid{⌉ᵁ}\;\mathrel{=}\;\Conid{Entity}{}\<[E]%
\\
\>[3]{}\Varid{⌈}\;\Conid{S}\;{}\<[9]%
\>[9]{}\Varid{⌉ᵁ}\;\mathrel{=}\;\Conid{Bool}{}\<[E]%
\ColumnHook
\end{hscode}\resethooks
Now that we have \ensuremath{\Conid{Univ}} and \ensuremath{\Varid{⌈\char95 ⌉ᵁ}}, we can open up the modules defined
as above, instantiating the return type \ensuremath{\Conid{R}} with the type of booleans.
\begin{hscode}\SaveRestoreHook
\column{B}{@{}>{\hspre}l<{\hspost}@{}}%
\column{3}{@{}>{\hspre}l<{\hspost}@{}}%
\column{27}{@{}>{\hspre}l<{\hspost}@{}}%
\column{E}{@{}>{\hspre}l<{\hspost}@{}}%
\>[3]{}\Keyword{open}\;\Varid{logic}\;{}\<[27]%
\>[27]{}\Conid{Univ}{}\<[E]%
\\
\>[3]{}\Keyword{open}\;\Varid{logic.translation}\;{}\<[27]%
\>[27]{}\Conid{Univ}\;\Varid{⌈\char95 ⌉ᵁ}\;\Conid{Bool}{}\<[E]%
\ColumnHook
\end{hscode}\resethooks
With everything that we implemented in scope, we can now define a
small lexicon for our example sentence.

In what follows, we will use the aliases \ensuremath{\Varid{n}}, \ensuremath{\Varid{np}} and \ensuremath{\Varid{s}} for \ensuremath{\Varid{el}\;\Conid{N}},
\ensuremath{\Varid{el}\;\Conid{NP}} and \ensuremath{\Varid{el}\;\Conid{S}}, respectively:
\begin{hscode}\SaveRestoreHook
\column{B}{@{}>{\hspre}l<{\hspost}@{}}%
\column{3}{@{}>{\hspre}l<{\hspost}@{}}%
\column{13}{@{}>{\hspre}l<{\hspost}@{}}%
\column{E}{@{}>{\hspre}l<{\hspost}@{}}%
\>[3]{}\Varid{someone}\;{}\<[13]%
\>[13]{}\mathbin{:}\;\Varid{⌈}\;(\Varid{np}\;\Varid{⇐}\;\Varid{n})\;\Varid{⊗}\;\Varid{n}\;\Varid{⌉}{}\<[E]%
\\
\>[3]{}\Varid{someone}\;{}\<[13]%
\>[13]{}\mathrel{=}\;((\Varid{λ}\;\{\mskip1.5mu (\Varid{g}\;\Varid{,}\;\Varid{f})\;\Varid{→}\;\exists\;(\Varid{λ}\;\Varid{x}\;\Varid{→}\;\Varid{f}\;\Varid{x}\;\Varid{∧}\;\Varid{g}\;\Varid{x})\mskip1.5mu\})\;\Varid{,}\;\textsc{person}){}\<[E]%
\\[\blanklineskip]%
\>[3]{}\Varid{loves}\;{}\<[13]%
\>[13]{}\mathbin{:}\;\Varid{⌈}\;(\Varid{np}\;\Varid{⇒}\;\Varid{s})\;\Varid{⇐}\;\Varid{np}\;\Varid{⌉}{}\<[E]%
\\
\>[3]{}\Varid{loves}\;{}\<[13]%
\>[13]{}\mathrel{=}\;\Varid{λ}\;\{\mskip1.5mu (\Varid{k}\;\Varid{,}\;\Varid{y})\;\Varid{→}\;\Varid{k}\;(\Varid{λ}\;\{\mskip1.5mu (\Varid{x}\;\Varid{,}\;\Varid{k})\;\Varid{→}\;\Varid{k}\;(\textsc{loves}\;\Varid{x}\;\Varid{y})\mskip1.5mu\})\mskip1.5mu\}{}\<[E]%
\\[\blanklineskip]%
\>[3]{}\Varid{everyone}\;{}\<[13]%
\>[13]{}\mathbin{:}\;\Varid{⌈}\;(\Varid{np}\;\Varid{⇐}\;\Varid{n})\;\Varid{⊗}\;\Varid{n}\;\Varid{⌉}{}\<[E]%
\\
\>[3]{}\Varid{everyone}\;{}\<[13]%
\>[13]{}\mathrel{=}\;((\Varid{λ}\;\{\mskip1.5mu (\Varid{g}\;\Varid{,}\;\Varid{f})\;\Varid{→}\;\forall\;(\Varid{λ}\;\Varid{x}\;\Varid{→}\;\Varid{f}\;\Varid{x}\;\Varid{⊃}\;\Varid{g}\;\Varid{x})\mskip1.5mu\})\;\Varid{,}\;\textsc{person}){}\<[E]%
\ColumnHook
\end{hscode}\resethooks
Given the types we used for our lexical entries, the judgement which
asserts the grammaticality of our sentence becomes:
\begin{hscode}\SaveRestoreHook
\column{B}{@{}>{\hspre}l<{\hspost}@{}}%
\column{5}{@{}>{\hspre}l<{\hspost}@{}}%
\column{E}{@{}>{\hspre}l<{\hspost}@{}}%
\>[5]{}((\Varid{np}\;\Varid{⇐}\;\Varid{n})\;\Varid{⊗}\;\Varid{n})\;\Varid{⊗}\;(((\Varid{np}\;\Varid{⇒}\;\Varid{s})\;\Varid{⇐}\;\Varid{np})\;\Varid{⊗}\;((\Varid{np}\;\Varid{⇐}\;\Varid{n})\;\Varid{⊗}\;\Varid{n}))\;\Varid{⊢}\;\Varid{s}{}\<[E]%
\ColumnHook
\end{hscode}\resethooks
There are seven proofs of this judgement. Below we have
included the first \textit{two} proofs:\footnote{%
  We have chosen not to include the other five proofs as, under the
  CPS translation, they have the same interpretations as either the
  first or the second proof.
  For the interested reader, however, the proofs are present in the
  source, and therefore available on GitHub.
}:
\setboolean{partial}{true}%
\begin{hscode}\SaveRestoreHook
\column{B}{@{}>{\hspre}l<{\hspost}@{}}%
\column{3}{@{}>{\hspre}l<{\hspost}@{}}%
\column{5}{@{}>{\hspre}l<{\hspost}@{}}%
\column{10}{@{}>{\hspre}l<{\hspost}@{}}%
\column{E}{@{}>{\hspre}l<{\hspost}@{}}%
\>[3]{}\textsc{sent}_{0}\;{}\<[10]%
\>[10]{}\mathrel{=}\;{}\<[E]%
\\
\>[3]{}\hsindent{2}{}\<[5]%
\>[5]{}\Varid{r⇒⊗}\;(\Varid{r⇐⊗}\;(\Varid{m⇐}\;(\Varid{m⇒}\;(\Varid{r⇐⊗}\;\Varid{ax′})\;\Varid{ax})\;(\Varid{r⇐⊗}\;\Varid{ax′}))){}\<[E]%
\\
\>[3]{}\textsc{sent}_{1}\;{}\<[10]%
\>[10]{}\mathrel{=}\;{}\<[E]%
\\
\>[3]{}\hsindent{2}{}\<[5]%
\>[5]{}\Varid{r⇐⊗}\;(\Varid{r⇐⊗}\;(\Varid{m⇐}\;(\Varid{r⊗⇐}\;(\Varid{r⇒⊗}\;(\Varid{r⇐⊗}\;(\Varid{m⇐}\;\Varid{ax′}\;(\Varid{r⇐⊗}\;\Varid{ax′})))))\;\Varid{ax})){}\<[E]%
\ColumnHook
\end{hscode}\resethooks
\setboolean{partial}{false}%
We can now apply our CPS translation to compute the denotations of our
sentence, passing in the denotations of the words as a tuple, and
passing in the identity function as the last argument in order to
obtain the result:
\setboolean{partial}{true}%
\begin{hscode}\SaveRestoreHook
\column{B}{@{}>{\hspre}l<{\hspost}@{}}%
\column{3}{@{}>{\hspre}l<{\hspost}@{}}%
\column{9}{@{}>{\hspre}l<{\hspost}@{}}%
\column{E}{@{}>{\hspre}l<{\hspost}@{}}%
\>[3]{}\Varid{sent₀}\;\mathrel{=}\;\Varid{⌈}\;\textsc{sent}_{0}\;\Varid{⌉ᴿ}\;(\Varid{someone}\;\Varid{,}\;\Varid{loves}\;\Varid{,}\;\Varid{everyone})\;\Varid{id}\;{}\<[E]%
\\
\>[3]{}\hsindent{6}{}\<[9]%
\>[9]{}\mapsto\;\forall\;(\Varid{λ}\;\Varid{y}\;\Varid{→}\;\textsc{person}\;\Varid{y}\;\Varid{⊃}\;\exists\;(\Varid{λ}\;\Varid{x}\;\Varid{→}\;\textsc{person}\;\Varid{x}\;\Varid{∧}\;\textsc{loves}\;\Varid{x}\;\Varid{y})))\;{}\<[E]%
\\
\>[3]{}\Varid{sent₁}\;\mathrel{=}\;\Varid{⌈}\;\textsc{sent}_{1}\;\Varid{⌉ᴿ}\;(\Varid{someone}\;\Varid{,}\;\Varid{loves}\;\Varid{,}\;\Varid{everyone})\;\Varid{id}\;{}\<[E]%
\\
\>[3]{}\hsindent{6}{}\<[9]%
\>[9]{}\mapsto\;\exists\;(\Varid{λ}\;\Varid{x}\;\Varid{→}\;\textsc{person}\;\Varid{x}\;\Varid{∧}\;\forall\;(\Varid{λ}\;\Varid{y}\;\Varid{→}\;\textsc{person}\;\Varid{y}\;\Varid{⊃}\;\textsc{loves}\;\Varid{x}\;\Varid{y})){}\<[E]%
\ColumnHook
\end{hscode}\resethooks
\setboolean{partial}{false}%
\textit{Voila}! Our system produces exactly the expected readings.

\begin{figure*}[ht]
\centering
\begin{minipage}[b]{0.495\textwidth}%
\centering%
\begin{scprooftree}{0.85}%
\RightLabel{\tiny{$(\text{ax})$}}%
\AXC{$np \fCenter np$}%
\RightLabel{\tiny{$(\text{ax})$}}%
\AXC{$n \fCenter n$}%
\RightLabel{\tiny{$(m\varslash)$}}%
\BIC{$np \varslash n \fCenter np \varslash n$}%
\RightLabel{\tiny{$(r\varslash\varotimes)$}}%
\UIC{$(np \varslash n) \varotimes n \fCenter np$}%
\RightLabel{\tiny{$(\text{ax})$}}%
\AXC{$s \fCenter s$}%
\RightLabel{\tiny{$(m\varbslash)$}}%
\BIC{$np \varbslash s \fCenter (np \varslash n) \varotimes n \varbslash s$}%
\RightLabel{\tiny{$(\text{ax})$}}%
\AXC{$np \fCenter np$}%
\RightLabel{\tiny{$(\text{ax})$}}%
\AXC{$n \fCenter n$}%
\RightLabel{\tiny{$(m\varslash)$}}%
\BIC{$np \varslash n \fCenter np \varslash n$}%
\RightLabel{\tiny{$(r\varslash\varotimes)$}}%
\UIC{$(np \varslash n) \varotimes n \fCenter np$}%
\RightLabel{\tiny{$(m\varslash)$}}%
\BIC{$(np \varbslash s) \varslash np \fCenter ((np \varslash n) \varotimes n \varbslash s) \varslash (np \varslash n) \varotimes n$}%
\RightLabel{\tiny{$(r\varslash\varotimes)$}}%
\UIC{$((np \varbslash s) \varslash np) \varotimes (np \varslash n) \varotimes n \fCenter (np \varslash n) \varotimes n \varbslash s$}%
\RightLabel{\tiny{$(r\varbslash\varotimes)$}}%
\UIC{$((np \varslash n) \varotimes n) \varotimes ((np \varbslash s) \varslash np) \varotimes (np \varslash n) \varotimes n \fCenter s$}%
\end{scprooftree}%
$\forall\,(\lambda y\,→\,\textsc{person}\,y\,\supset\,\exists\,(\lambda x\,→\,\textsc{person}\,x\,\land\,\textsc{loves}\,x\,y))$%
\end{minipage}%
\begin{minipage}[b]{0.495\textwidth}%
\centering%
\begin{scprooftree}{0.85}%
\RightLabel{\tiny{$(\text{ax})$}}%
\AXC{$np \fCenter np$}%
\RightLabel{\tiny{$(\text{ax})$}}%
\AXC{$s \fCenter s$}%
\RightLabel{\tiny{$(m\varbslash)$}}%
\BIC{$np \varbslash s \fCenter np \varbslash s$}%
\RightLabel{\tiny{$(\text{ax})$}}%
\AXC{$np \fCenter np$}%
\RightLabel{\tiny{$(\text{ax})$}}%
\AXC{$n \fCenter n$}%
\RightLabel{\tiny{$(m\varslash)$}}%
\BIC{$np \varslash n \fCenter np \varslash n$}%
\RightLabel{\tiny{$(r\varslash\varotimes)$}}%
\UIC{$(np \varslash n) \varotimes n \fCenter np$}%
\RightLabel{\tiny{$(m\varslash)$}}%
\BIC{$(np \varbslash s) \varslash np \fCenter (np \varbslash s) \varslash (np \varslash n) \varotimes n$}%
\RightLabel{\tiny{$(r\varslash\varotimes)$}}%
\UIC{$((np \varbslash s) \varslash np) \varotimes (np \varslash n) \varotimes n \fCenter np \varbslash s$}%
\RightLabel{\tiny{$(r\varbslash\varotimes)$}}%
\UIC{$np \varotimes ((np \varbslash s) \varslash np) \varotimes (np \varslash n) \varotimes n \fCenter s$}%
\RightLabel{\tiny{$(r\varotimes\varslash)$}}%
\UIC{$np \fCenter s \varslash ((np \varbslash s) \varslash np) \varotimes (np \varslash n) \varotimes n$}%
\RightLabel{\tiny{$(\text{ax})$}}%
\AXC{$n \fCenter n$}%
\RightLabel{\tiny{$(m\varslash)$}}%
\BIC{$np \varslash n \fCenter (s \varslash ((np \varbslash s) \varslash np) \varotimes (np \varslash n) \varotimes n) \varslash n$}%
\RightLabel{\tiny{$(r\varslash\varotimes)$}}%
\UIC{$(np \varslash n) \varotimes n \fCenter s \varslash ((np \varbslash s) \varslash np) \varotimes (np \varslash n) \varotimes n$}%
\RightLabel{\tiny{$(r\varslash\varotimes)$}}%
\UIC{$((np \varslash n) \varotimes n) \varotimes ((np \varbslash s) \varslash np) \varotimes (np \varslash n) \varotimes n \fCenter s$}%
\end{scprooftree}%
$\exists\,(\lambda x\,\to\,\textsc{person}\,x\,\land\,\forall\,(\lambda y\,\to\,\textsc{person}\,y\,\supset\,\textsc{loves}\,x\,y))$%
\end{minipage}%
\caption{``Someone loves everyone.''}\label{someone_loves_everyone}%
\end{figure*}

\section{Conclusion}
We have presented the reader with a simple formalisation of the
Lambek-Grishin calculus, using the proof assistant Agda. We have shown
how to formalise the proof of the admissibility of cut from
\citet{moortgat1999} in Agda, and have extended this proof to cover
all of LG. While we have not covered any of the usual unary operators,
the formalism presented here generalises straightforwardly to
accommodate connectives of any arity---and this extension, together
with many other extensions, are indeed implemented in the full version
of our code.

We have then presented the reader with a call-by-value CPS
translation into the host language Agda, and used this translation to
demonstrate the analysis of an example sentence.

Most importantly, we hope we presented the reader with a clean and
readable formalisation of the Lambek-Grishin calculus.

\section{Related Work}
Previous work on the formalisation of Lambek calculi was done in Coq by \citet{anoun2004}.

The work presented in this paper is part of a larger undertaking to
formalise type-logical grammars in Agda. At the moment, we have
formalised not only the algebraic Lambek-Grishin calculus---which was
presented in this paper---but also structural and polarised varieties
of this calculus. From these implementations, we are able to extract
implementations of their respective non-associative Lambek calculi.

In addition, we have implemented various other multi-modal systems,
such as $\text{NL}_{\textit{CL}}$~\citep{bs2015}.

We aim to extend this work by further formalising the known work
w.r.t.\ these calculi, and creating tools to accommodate the writing
of formal linguistics papers in literate style.

\nocite{*}
\bibliographystyle{apalike}
\bibliography{main}

\end{document}